\documentclass{desyproc}
\usepackage{hyperref}

\begin{document}


\title{Obtaining the CMS Ridge effect with Multiparton Interactions}



%
%
%
%
%
%
%
%
%

\author{{\slshape Sara Alderweireldt$^1$, Pierre Van Mechelen$^1$}\\[1ex]
$^1$Universiteit Antwerpen, Groenenborgerlaan 171, 2020 Antwerpen, Belgi\"e}


\hypersetup{
  pdfauthor={Sara Alderweireldt, Pierre Van Mechelen},
  pdfsubject={CMS ridge},
  pdftitle={Obtaining the CMS Ridge effect with Multiparton Interactions},
  pdfkeywords={CMS ridge, monte carlo, multiparton interactions, tuning},
  pdfcreator={pdflatex},
  pdfproducer={Sara Alderweireldt},
  pdfborder=0 0 0
}

\contribID{ZZ}
\confID{UU}
\desyproc{DESY-PROC-2012-YY}
\acronym{MPI@LHC 2011}
\doi
\maketitle


\begin{abstract}
We show that the ``ridge'' phenomenon in the two-particle angular correlation function, as observed by the CMS experiment, can be reproduced by implementing an impact parameter dependent azimuthal correlation of the scattering planes of individual partonic interactions.  Such an approach is motivated by the observation that even for moderate impact parameters a substantial number of partonic interactions may be produced, while at the same time the protons are sufficiently far apart to create a preferential direction in azimuth.

A re-tune of the \textsc{Pythia6} Z2 tune based on underlying event and minimum bias distributions measured at the LHC shows that a better description of data can be obtained with this approach and that some tension existing between underlying event and minimum bias distributions can be removed. We show that, even though the CMS result on the angular correlation function itself is not used in the re-tune, we can predict the appearance of long-range, near-side angular correlations in proton-proton collisions.

\end{abstract}



\section{Introduction}
The CMS ridge effect is a two-particle angular correlation effect observed by the Compact Muon Solenoid (CMS) experiment in Large Hadron Collider (LHC) proton-proton collisions at high charged track multiplicities ($N_{ch} > 110$) and in a specific transverse momentum range ($p_T = 1-3$~GeV). This effect, together with other visible structures in the $R\left(\Delta\eta,\Delta\phi\right)$ correlation distribution, was described in \cite{cmsridge}. \vskip 1em \noindent
Two-particle correlation function $R\left(\Delta\eta,\Delta\phi\right)$ is defined as:
\begin{equation}
R\left(\Delta\eta,\Delta\phi\right) = \left\langle \left(\langle N\rangle - 1 \right) \left( \frac{S_N\left(\Delta\eta,\Delta\phi\right)}{B_N\left(\Delta\eta,\Delta\phi\right)}-1\right)\right\rangle_{bins}
\end{equation}
Data is binned according to charged track multiplicity $N_{ch}$. The signal $S_N(\Delta\eta,\Delta\phi)$ consists of the charged two-particle density, while the background $B_N(\Delta\eta,\Delta\phi)$ is given by the distribution of uncorrelated particle pairs -- the product of two single-particle distributions. Finally, the data is averaged, weighted with bin multiplicity, over all bins. The analysis is repeated for four sets of data. On one hand two minimum bias sets (all $N_{ch}$), one including all particles with transverse momenta above $0.1$~GeV and the other including all particles with transverse momenta between $1$ and $3$~GeV. On the other hand two high-multiplicity sets ($N_{ch} > 110$), again with the same two transerverse momentum selections. \vskip 1em \noindent
Some of the effects reported in \cite{cmsridge}, including the near-side peak at $\left(\Delta\eta,\Delta\phi\right) = \left(0,0\right)$, the away-side ridge at $\left(\Delta\eta,\Delta\phi\right) = \left(\Delta\eta,\pi\right)$ and the Gaussian ridge at $\left(\Delta\eta,\Delta\phi\right) = \left(0,\Delta\phi\right)$ can be explained with single two-to-two partonic interactions. The first two are visible in all sets of data, while the third one is most clear in the minimum bias  $p_T > 0.1$~GeV case. A fourth effect, the near-side ridge, a long-range azimuthal correlation at $\left(\Delta\eta,\Delta\phi\right) = \left(\Delta\eta,0\right)$ only visible in high-multiplicity events at moderate $p_T$, requires further study. It is for this last effect that we propose a model. \vskip 1em \noindent
For our study we observe the effect of our modification of the \textsc{Pythia6}~\cite{pythia6} Monte Carlo (MC) event generator on select observables and consider changes in a few existing \textsc{Pythia6} parameters to counteract the side-effects of our modification. This latter step can be considered a re-tuning to CMS data. Note that we only use a limited set of CMS data and start from the existing \textsc{Pythia6} tune Z2~\cite{cmsue}. More global tuning including other experiments' data was not within the scope of this study, but may be added later.

\section{The azimuthal alignment model}
\begin{figure}[h]
\centering
\includegraphics[width=0.35\textwidth]{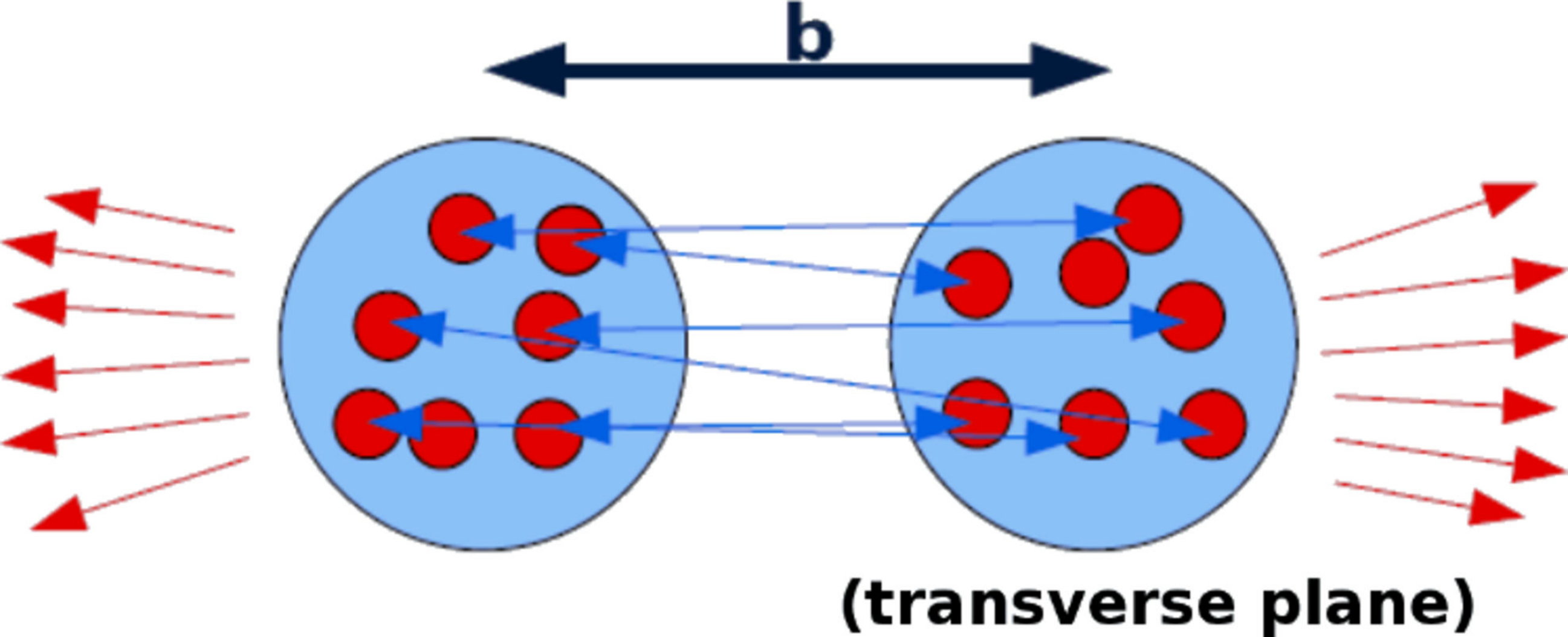} 
\caption{\label{fig_pprofile}Protons separated by impact parameter $b$.}
\end{figure}

\begin{figure}[h]
\centering
\includegraphics[width=0.31\textwidth]{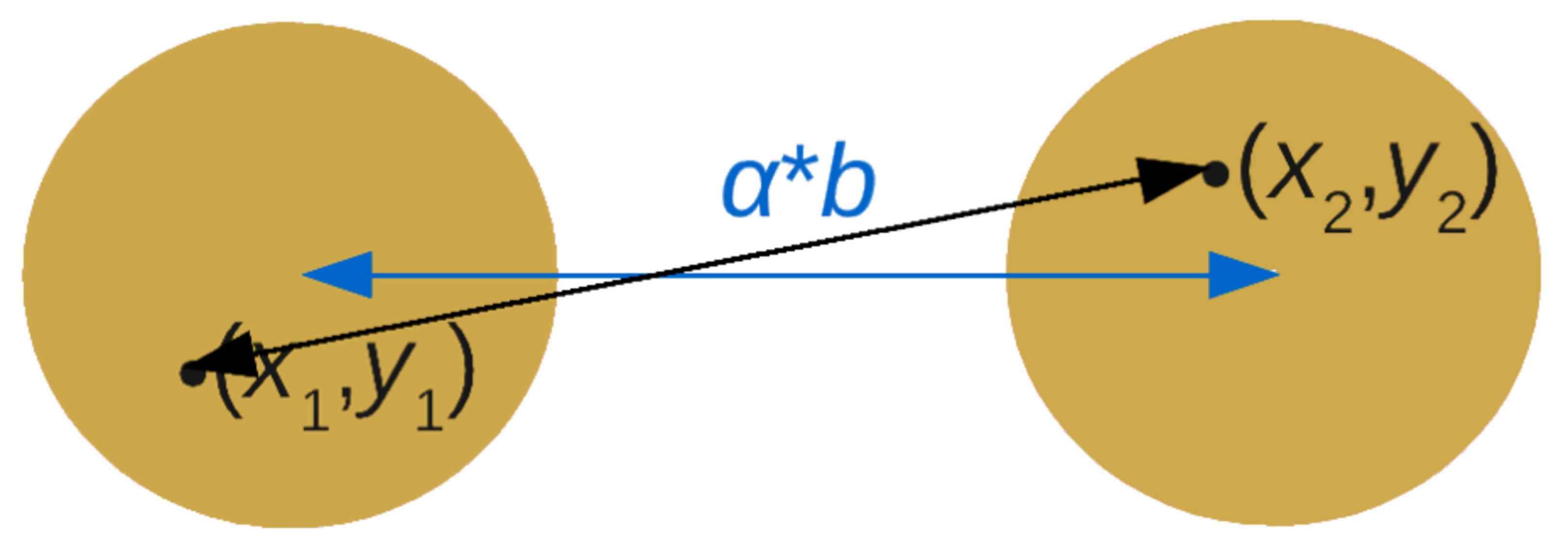} 
\caption{\label{fig_sample}Sampling of random points in Gaussian proton profiles, separated by impact parameter $b$, and introduction of scaling parameter $\alpha$ to allow some tuning freedom.}
\end{figure}

\begin{figure}[!h]
\centering
\includegraphics[width=0.25\textwidth]{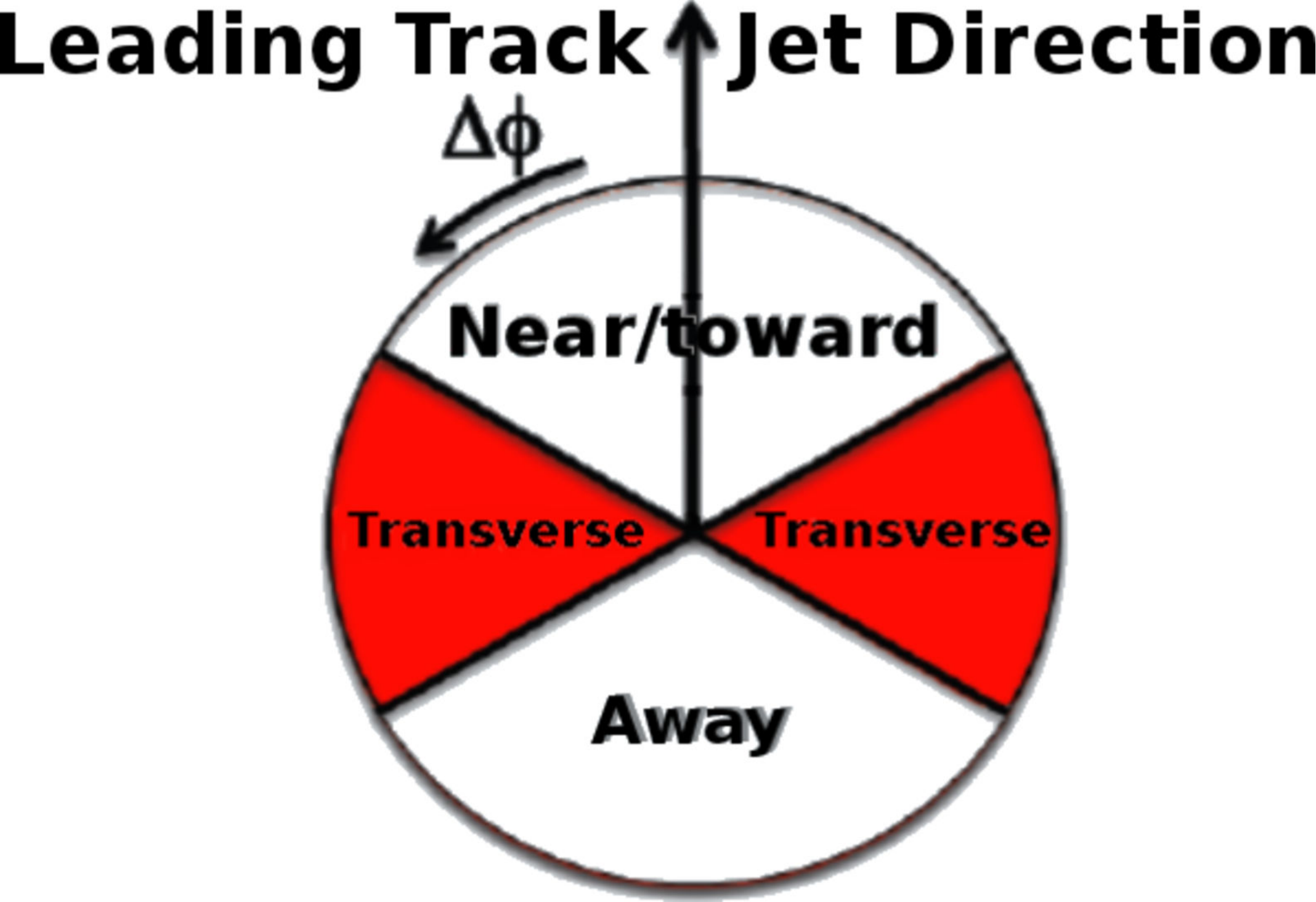} 
\caption{\label{fig_tta}Areas in $\Delta\phi$ with respect to the leading track jet.}
\end{figure}

\begin{figure}[ht]
\centering
\vspace{-0.5em}
\hfill\includegraphics[width=0.302\textwidth]{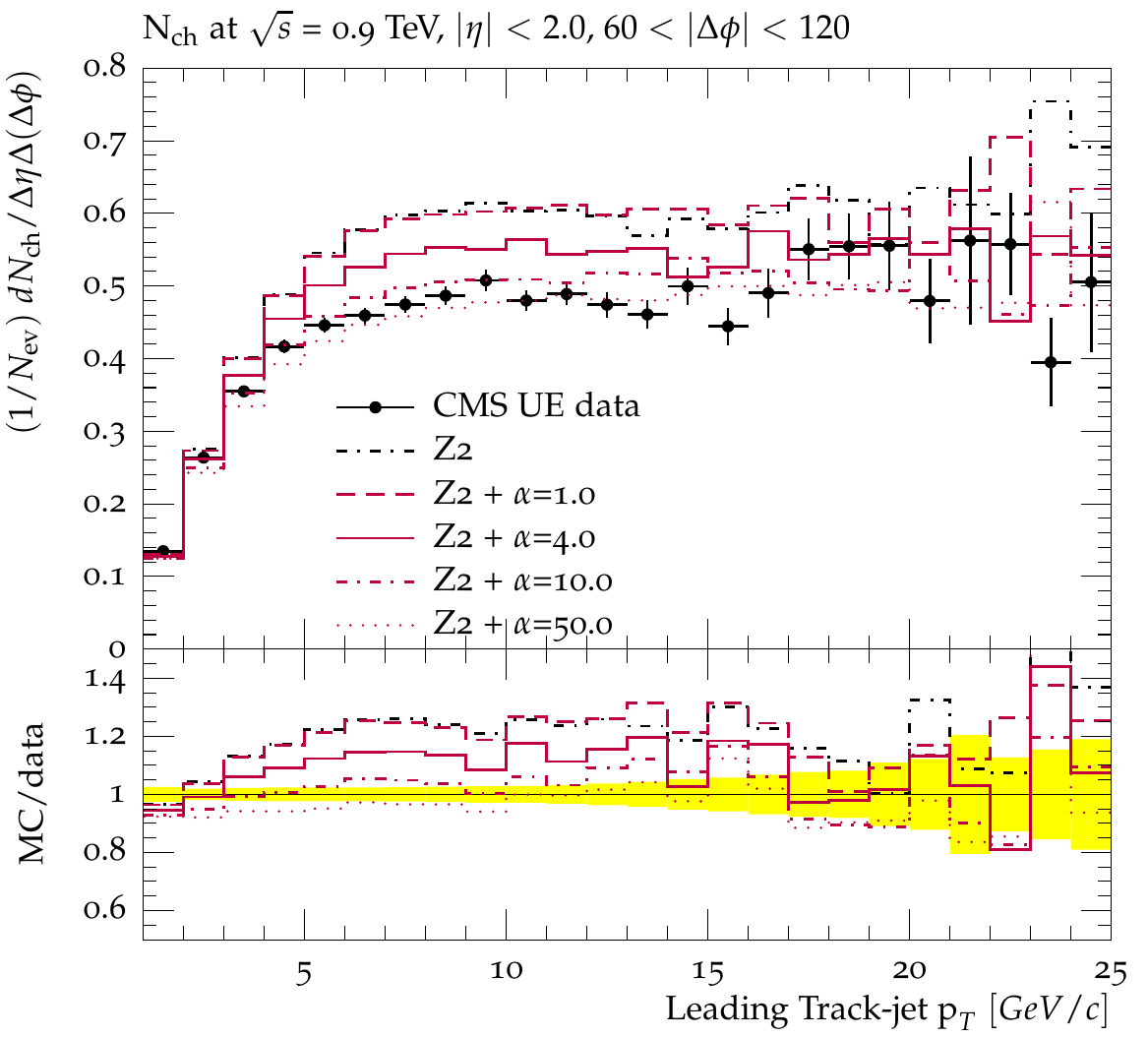}\hspace{1cm}
\includegraphics[width=0.302\textwidth]{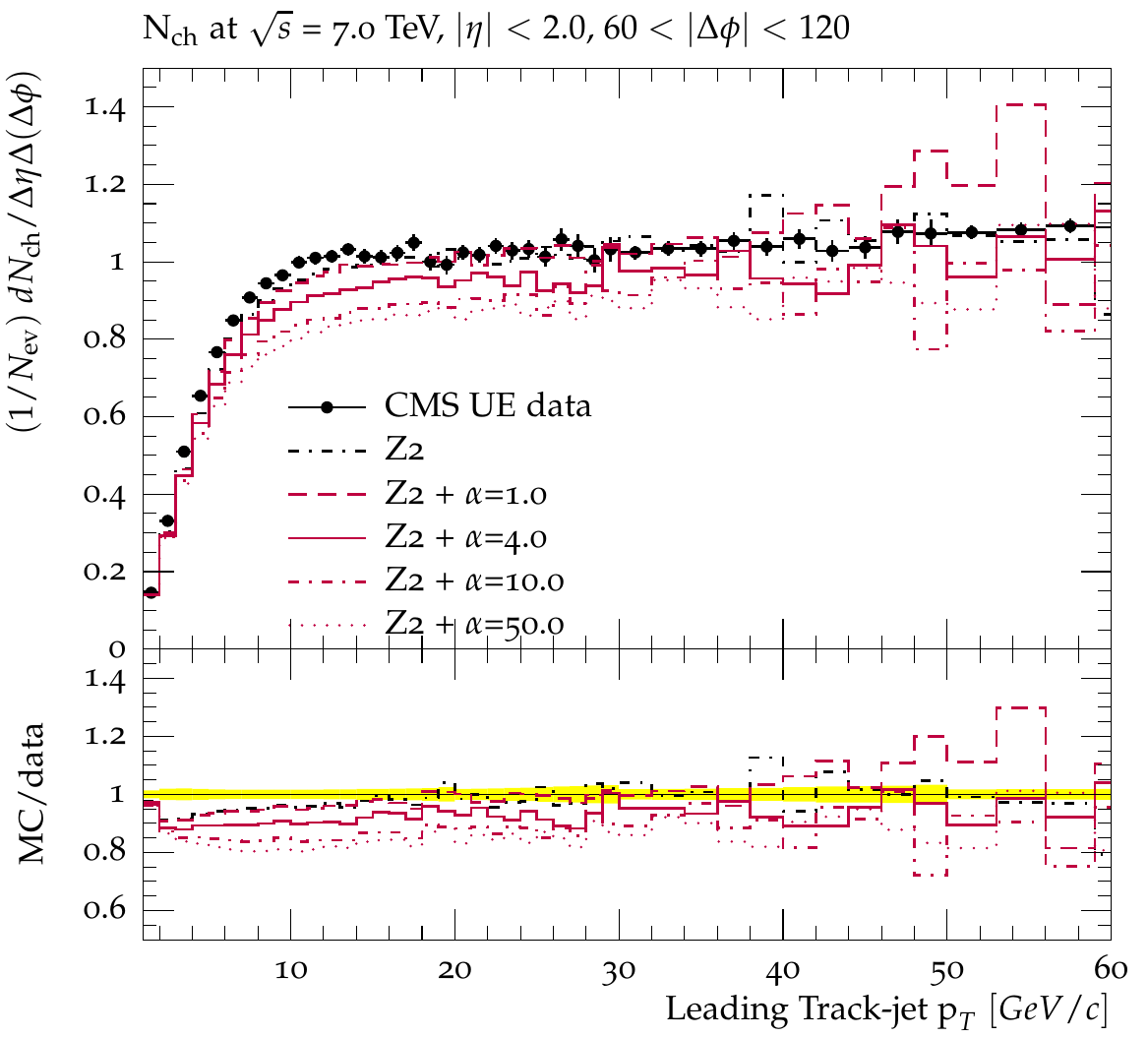}\hfill\mbox{\ }\vskip 0em
\hfill\includegraphics[width=0.302\textwidth]{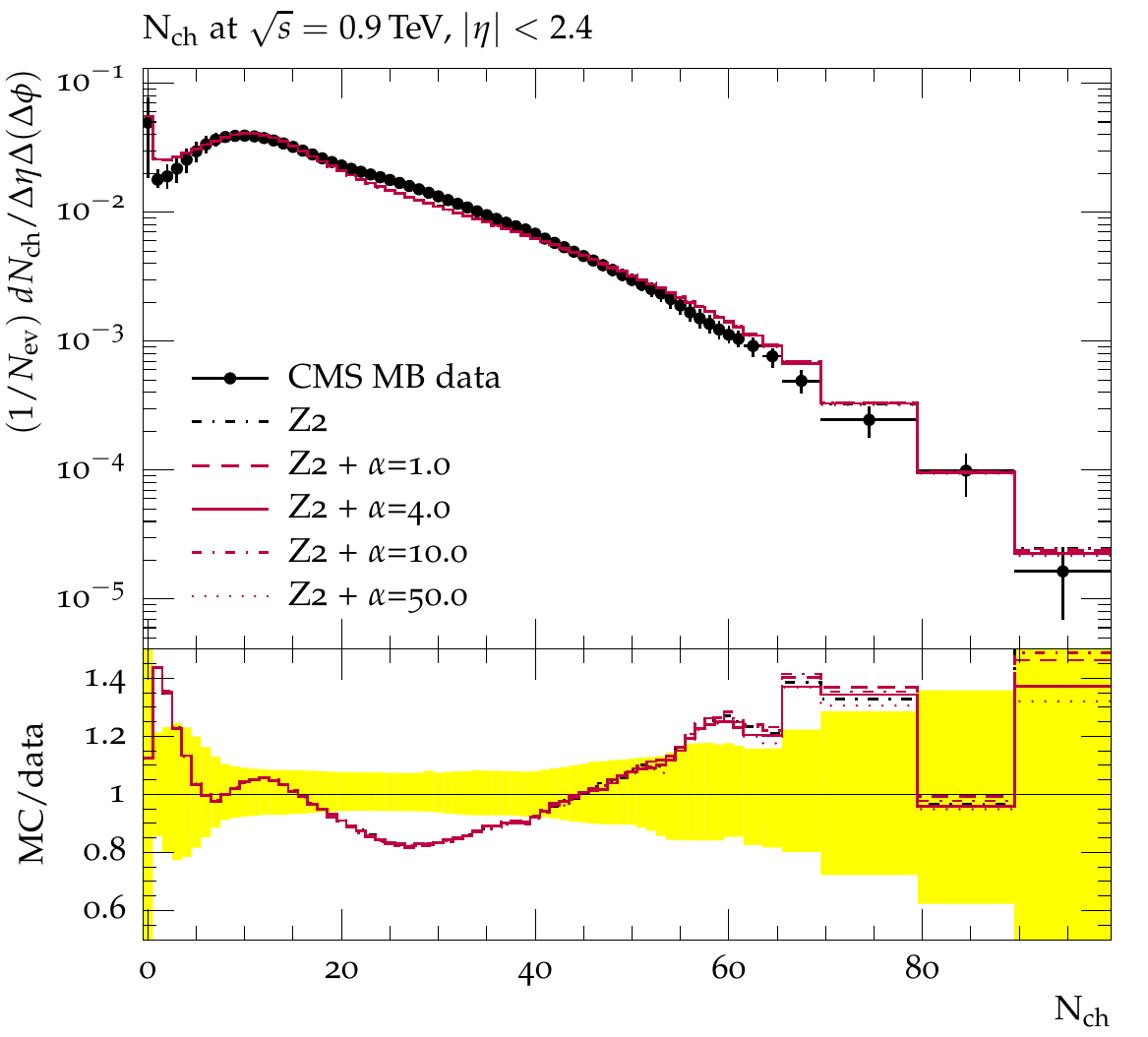}\hspace{1cm}
\includegraphics[width=0.302\textwidth]{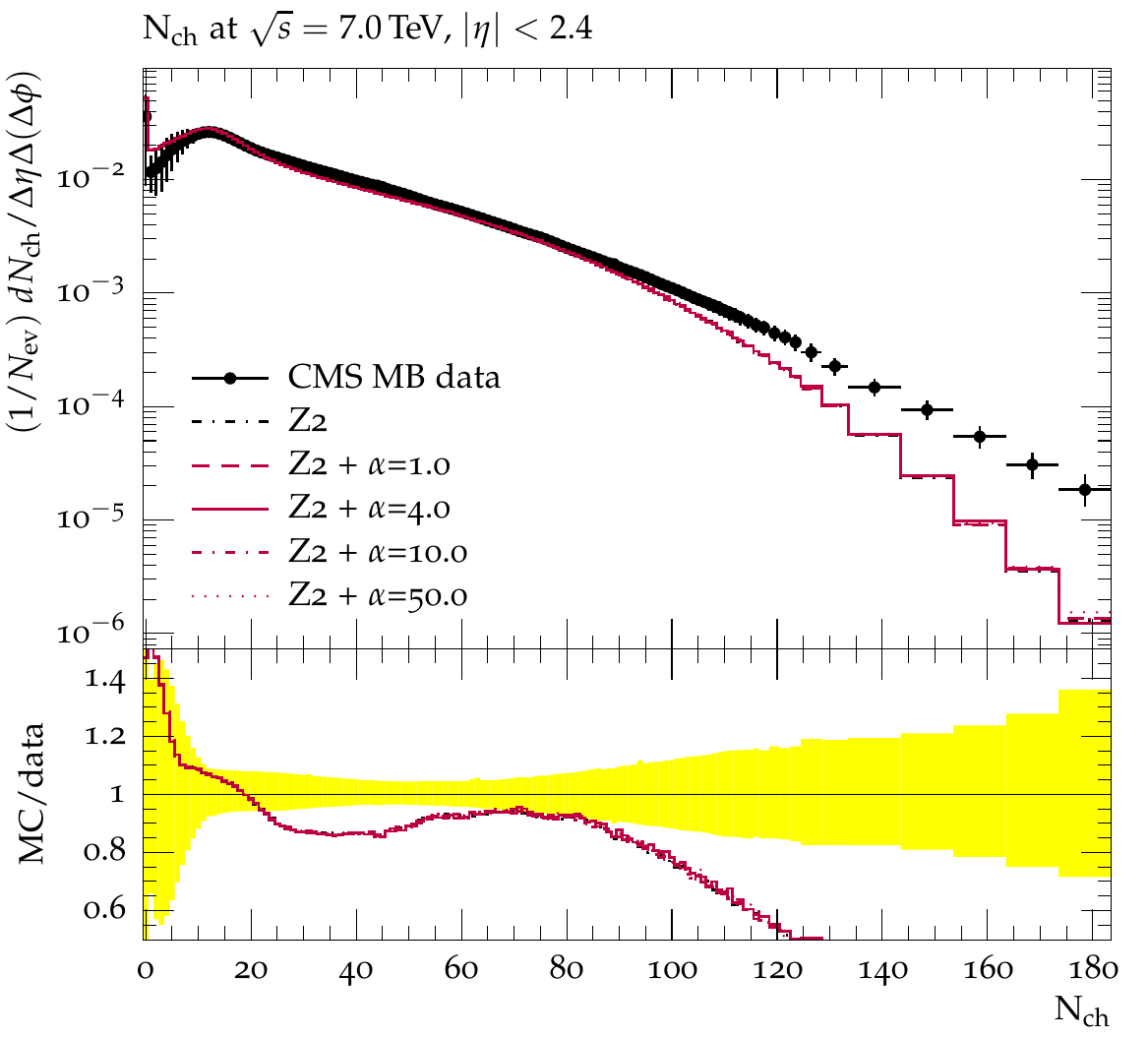}\hfill\mbox{\ }\vspace{-0.8em}
\caption{\label{fig_sens2}Overview of the sensitivity of $N_{ch}$ observables in CMS UE (top) and MB (bottom) data to changes in scaling parameter \textcolor{magenta}{$\alpha$ [purple]}, with Z2 [black] as a reference.}
\end{figure}

For large enough impact parameter $b$ (figure \ref{fig_pprofile}), the multiparton interactions in proton-proton collisions tend to lie in the collision plane of the hardest interaction and the final state particles will have similar azimuthal angle $\phi$ -- this results in near-side effects. Furthermore, an explanation for the ridge effect with multiparton interactions would require enough such interactions to be taking place, which leads to high-multiplicity events. At the same time we require that the multiparton interactions are semi-hard, and thus yield moderate-$p_T$ particles. Finally, we are dealing with incoming partons with very different $x_{bj}$ and as such will have interactions in a broad pseudo-rapidity range $\eta$ -- this gives rise to long-range effects. So far, everything is still consistent with the observations made by CMS.
\vskip 0em \noindent
What is still a problem, is that high-multiplicity events are generally central collisions which have an impact parameter $b \sim 0$, while the definition of the collision plane of the hardest interaction requires large $b$. In light of this issue, we study whether a small upward fluctuation in the amount of multiparton interactions, for the case of moderate impact parameter, suffices to explain the CMS ridge effect. \vskip 1em \noindent
The modification we introduce goes on top of the most recent multiparton interaction model currently in \textsc{Pythia6}~\cite{pythia6mpi}. In this existing model, the amount of multiparton interactions, a measure for the activity, is inversely proportional to impact parameter $b$ (VINT(139), rescaled to $b_{avg}=1$ for the minimum bias case). The azimuthal angle $\hat{\phi}$ (VINT(24)) is chosen randomly. This last point makes that angular correlations -- also the long-range, near-side ones -- would be missing in events generated with \textsc{Pythia6}. \vskip 1em \noindent 
We propose sampling random points $\left(x_i,y_i\right)$ in Gaussian proton profiles (figure \ref{fig_sample}), these protons being separated by impact parameter $b$, and using trigonometry to calculate the $\phi$-offset from the hardest interaction. To allow for some tuning freedom we add a scaling parameter $\alpha$ to the impact parameter $b$. Ideally, the scaling parameter would be one. This results in:
\begin{equation}
\phi_i = \phi_{hardest} + \mbox{arctan}\left(\frac{y_2-y_1}{(x_2+\alpha\cdot b/b_{avg}) - x_1}\right)
\end{equation}
We implement the modification for two different modes of the multiparton interaction model of \textsc{Pythia6} which both make use of hadronic overlap according to Gaussian distributions. In those cases, the above $\phi$-definition makes sense. In our tuning activity reported in section \ref{sec_tuning}, we focus on the mode which uses double-gaussian matter profiles (MSTP(82) = 4). 
\vskip 0em \noindent
The modification has several implications. We study two sets of data: CMS underlying event (UE) data~\cite{cmsue}, showing the charged multiplicity $N_{ch}$ and transverse momentum sum $\sum p_T$ in the region transverse to a jet or hard interaction (figure \ref{fig_tta}), and CMS minimum bias (MB) data~\cite{cmsmb}, showing the charged multiplicity $N_{ch}$ integrated over azimuthal angle $\phi$. By introducing the modification, we generate interactions with an azimuthal separation from the hardest interaction smaller than would be the case with the previous uniform azimuthal distribution. The interactions get shifted to the toward/away regions and the plateau for $N_{ch}^{transverse}$ drops (figure \ref{fig_sens2}, top). Re-raising this plateau to describe the data requires a re-tune, modifying the $p_T$-cutoff and by proxy the activity, $N_{ch}$. The $p_T$-cutoff in \textsc{Pythia6} is given by:
\begin{equation}
p_T^{min}\left(E_{CM}\right) = p_T^0 \cdot \left(\frac{E_{CM}}{E_{REF}}\right)^{\gamma} = PARP(82)\cdot \left(\frac{E_{CM}}{E_{REF}}\right)^{PARP(90)}
\end{equation}
\vskip 1em \noindent 
In contrast with the clear effect on UE results, we expect little or no sensitivity to the modification for the MB results, which are integrated over azimuth $\phi$ (figure \ref{fig_sens2}, bottom). Possibly this diffence in sensitivity also allows to lift some of the tension which exists between the UE and MB descriptions. 

\section{Tuning}\label{sec_tuning}
In this second part we report the results of our small-scale automized three-parameter re-tuning to the two earlier described CMS data sets. We start off with a review of the sensitivity of the observables to the three parameters and end with two tunes, one simple tune to just four $N_{ch}$ distributions (transverse \& total $N_{ch}$ and 0.9 \& 7.0~TeV) and one two-step tune to all MB and UE observables in the data sets. The first tune allows us to get a feeling of the parameter space in play, while the second one aims to reach a more solid result fixing the $p_T$-cutoff based on the MB observables (insensitive to scaling parameter $\alpha$) and then using the UE observables to fix $\alpha$. For the actual tuning, we make use of the \textsc{Professor} package~\cite{prof}, which takes care of automated tuning based on \textsc{Rivet} plots~\cite{rivet} of observables with reference data. For the interpretation of the tune result we make the comparison with existing tune Z2* 
 rather than with tune Z2. Z2* is a \textsc{Professor} re-tuning of Z2 to CMS UE data, for parameters PARP(82) and PARP(90). 

\subsection{Sensitivity}\label{tuneSens}
The first observation (figure~\ref{fig_sens1}) is that \textsc{Pythia6} $p_T^0$-reference PARP(82) and energy-scaling parameter PARP(90) affect all activity, both transverse and total activity at both $0.9$ and $7.0$~TeV. For variations in the $p_T^0$-reference (red/blue solid) the effect is the same at both energies, while for variations in the energy-scaling parameter (green/orange dashed) the effect is opposite at the two energies. This is to be expected since $0.9$ and $7.0$~TeV lie on both sides of reference energy $1.8$~TeV used in the $p_T$-cutoff formula. On the part of $\alpha$ (figure~\ref{fig_sens2}), we can see a clear effect in the transverse region (UE dataset) and little to no effect in the $\phi$-integrated case (MB dataset). 
\definecolor{orange}{rgb}{1,0.5,0}
\begin{figure}[ht]
\centering
\hfill\includegraphics[width=0.33\textwidth]{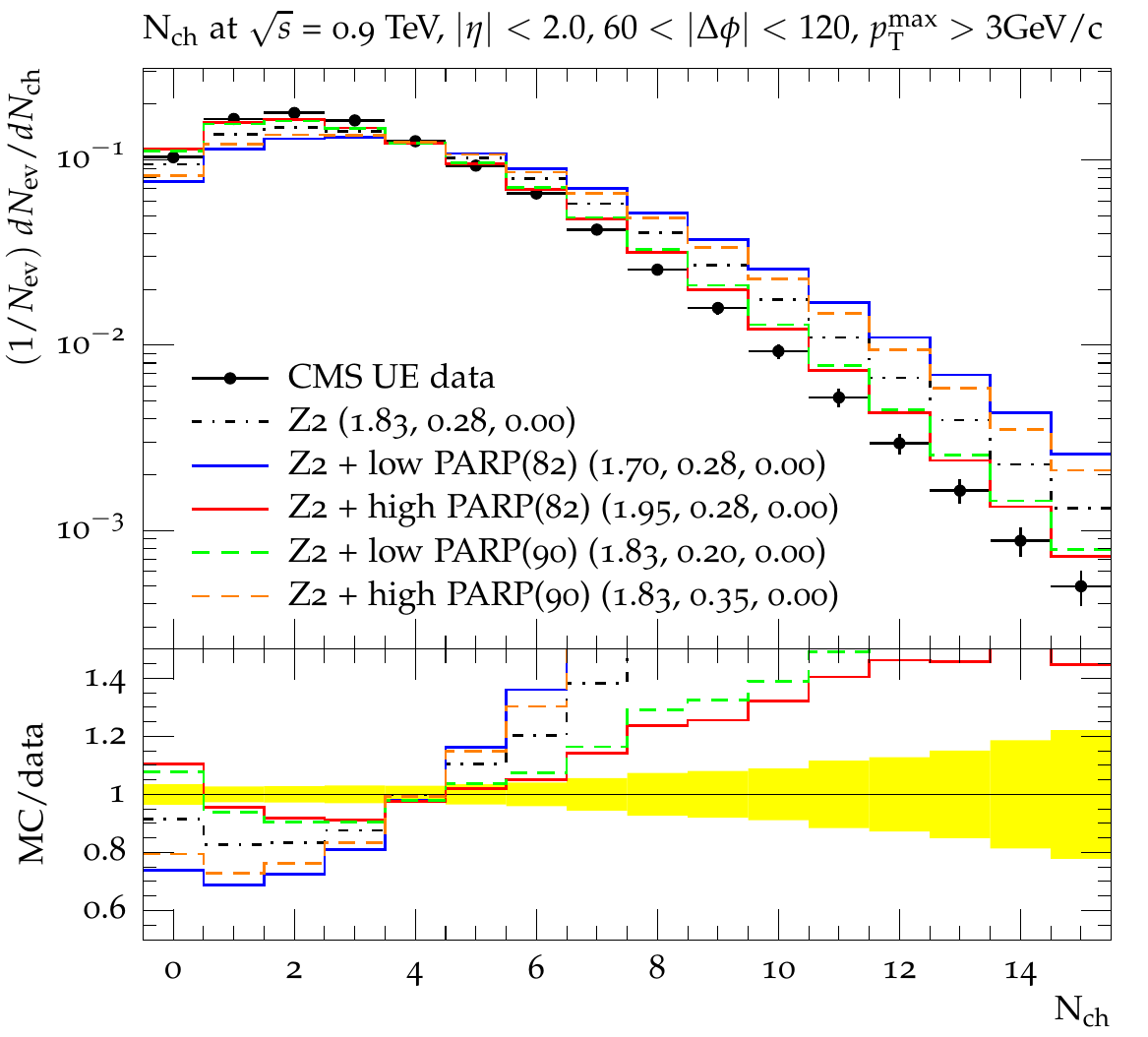}\hspace{1cm}
\includegraphics[width=0.33\textwidth]{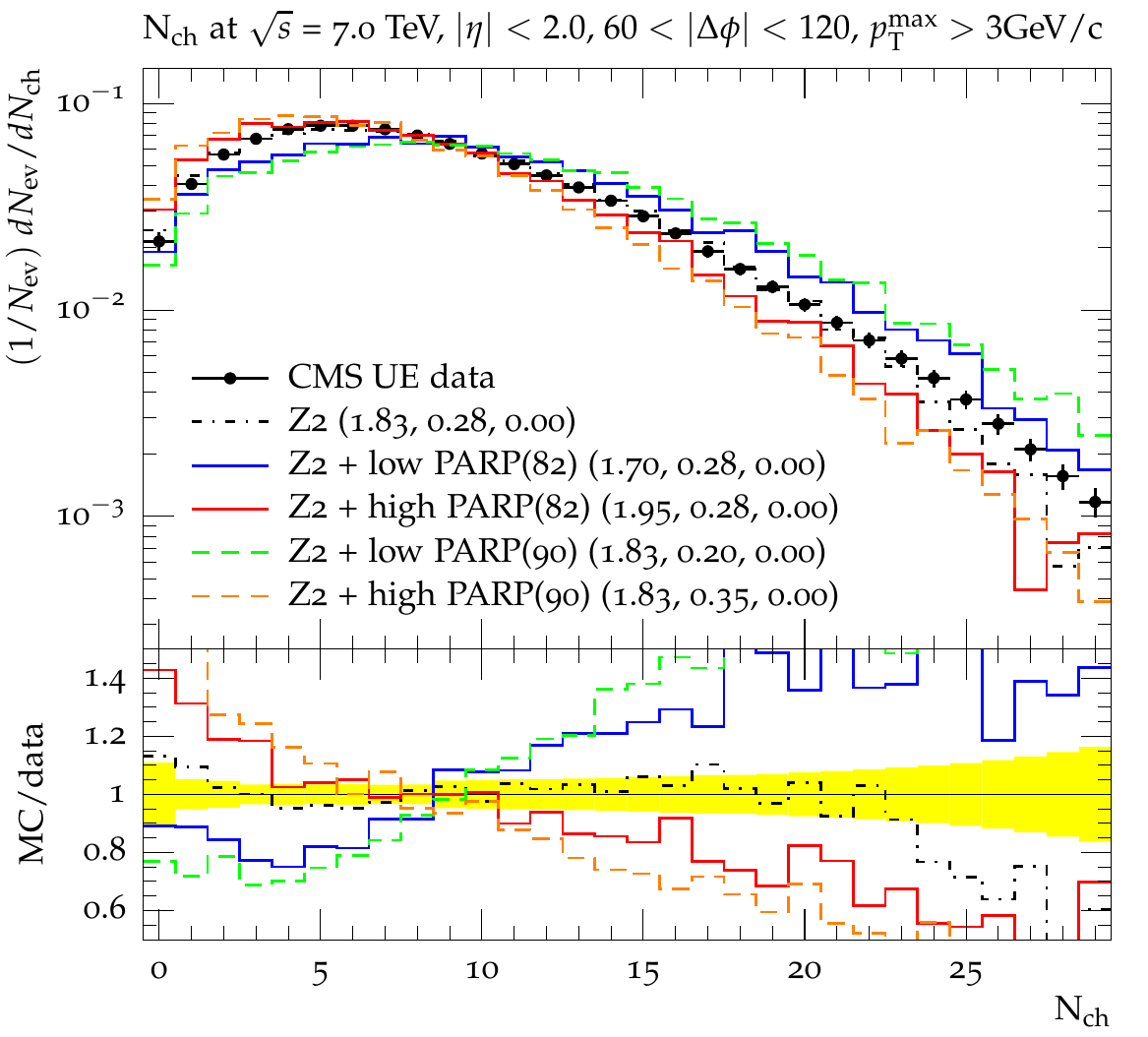}\hfill\mbox{\ }\vskip 0.2em
\hfill\includegraphics[width=0.33\textwidth]{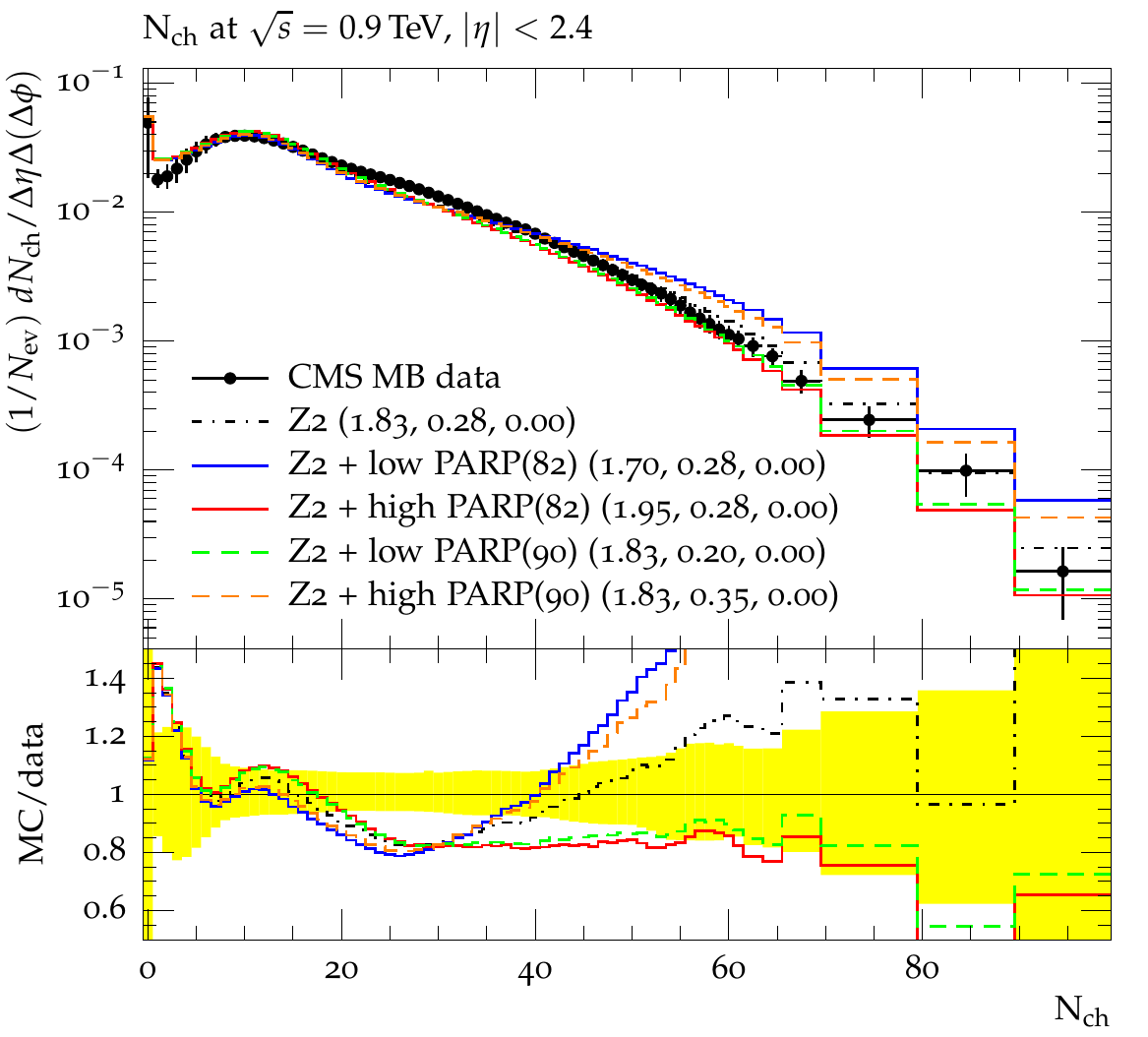}\hspace{1cm}
\includegraphics[width=0.33\textwidth]{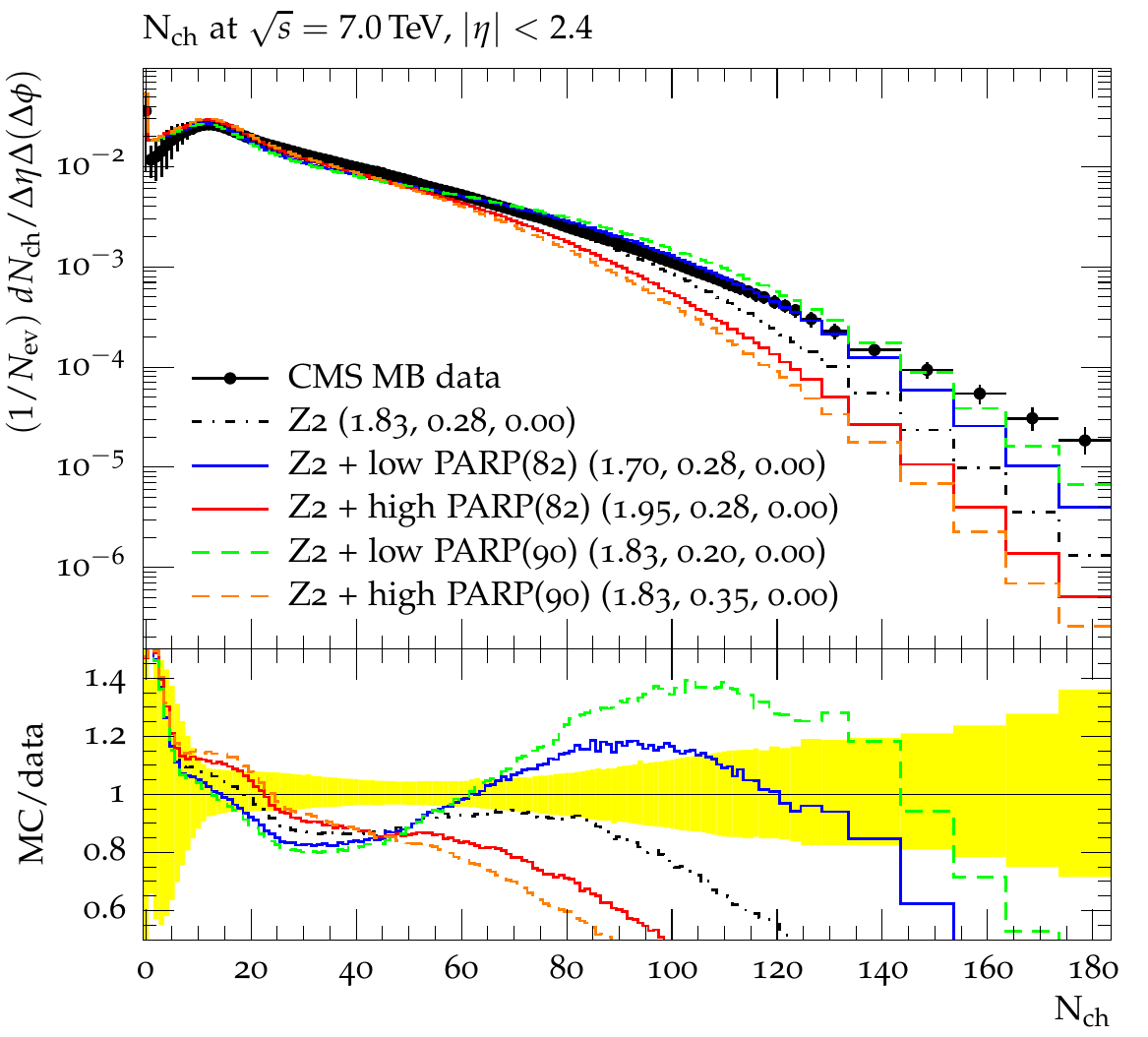}\hfill\mbox{\ }
\caption{\label{fig_sens1}Overview of the sensitivity of $N_{ch}$ observables in CMS UE (top) and MB (bottom) data to changes in PARP(82) [\textcolor{blue}{blue}, \textcolor{red}{red} solid] and PARP(90) [\textcolor{green}{green}, \textcolor{orange}{orange} dashed], with Z2 [black] as a reference. In brackets the values (PARP(82), PARP(90), $\alpha$) are marked.}
\end{figure}

\subsection{Tune one}\label{tuneOne}
This first crude tune, which we call Z2R, is made with the \textsc{Professor} package to just four observables (two UE and two MB) and confirms what is also more or less visible by eye. We want a moderate $\alpha$, so as not to destroy the match with data so much that it cannot be restored (either in MB, UE or both), while still having enough power to introduce the intended long-range near-side effect. Next, we need a slight lowering in the $p_T^0$-reference (compared to optimal tune Z2*) to re-raise the $N_{ch}$ plateau in the transverse region. The energy-dependence will be of less importance. We find exactly this in our Z2R \textsc{Professor} tune (table \ref{tab_Z2R}), for which we used the cubic interpolation mode. In general we find that, to begin with, the match with data for the four observables to which we tuned, is of the same quality as in the case of Z2*. Furthermore, also for the observables in the dataset which we did not include in the tuning, the match remains acceptable. We show the graphical result of tune Z2R (red solid) in figure \ref{fig_Z2RZ2Rp}, with tunes Z2 (black dotted) and Z2* (blue dashed) as reference. 

\begin{table}[ht]
\centering\footnotesize
\begin{tabular}{|c|c|c|c|}
\hline
& PARP(82) & PARP(90) & $\alpha$ \\ \hline\hline
Z2 & 1.83 & 0.28 & 0.00 \\ \hline
Z2* & 1.93 & 0.23 & 0.00 \\ \hline\hline
Z2R & 1.87 & 0.23 & 4.15 \\ \hline
\end{tabular}
\caption{Result of the 1-step 3-parameter tune to 4 observables.}
\label{tab_Z2R}
\end{table}

\subsection{Tune two}\label{tuneTwo}
\begin{figure}[p]
\centering
\hfill\includegraphics[width=0.33\textwidth]{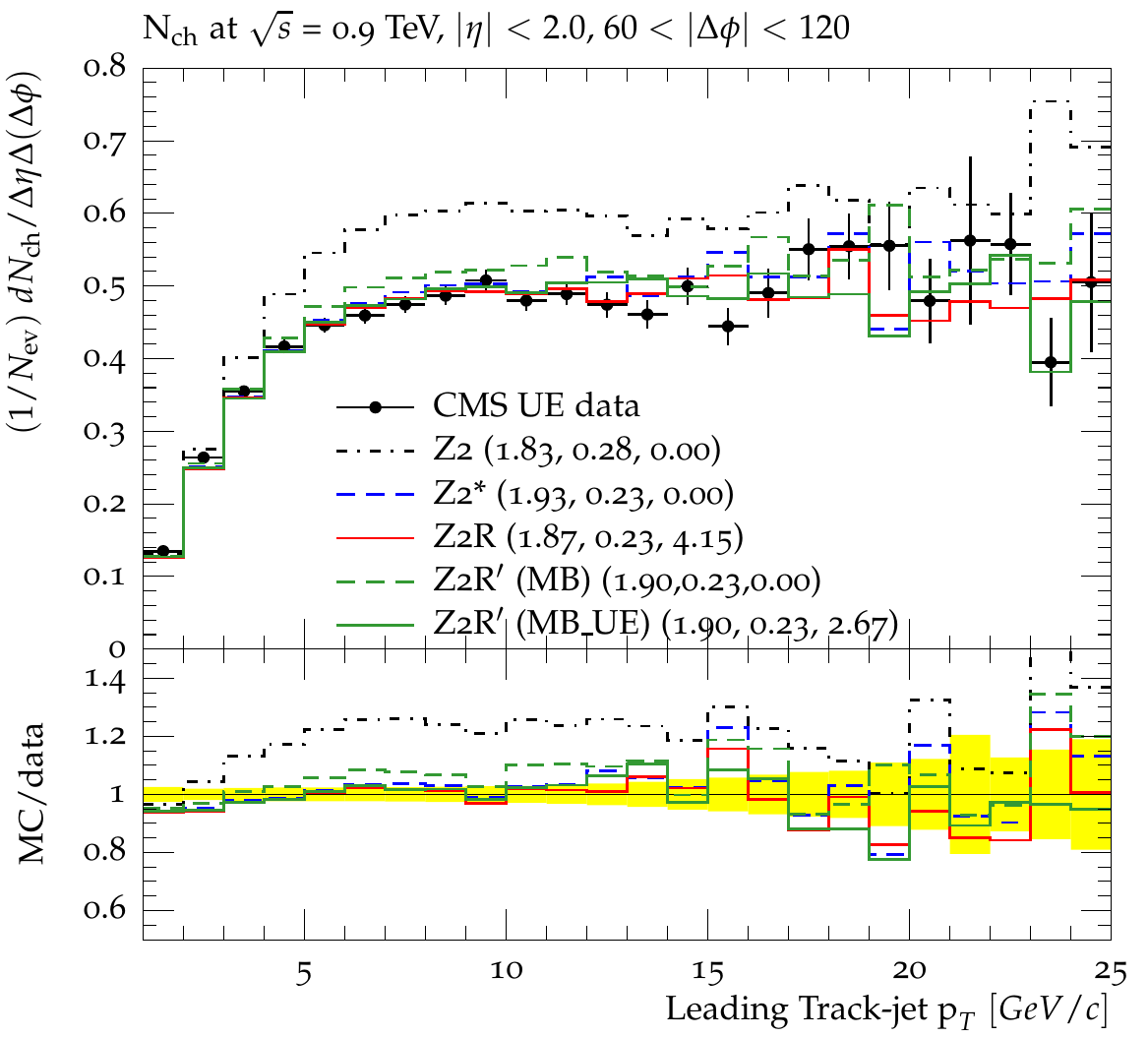}\hspace{1cm}
\includegraphics[width=0.33\textwidth]{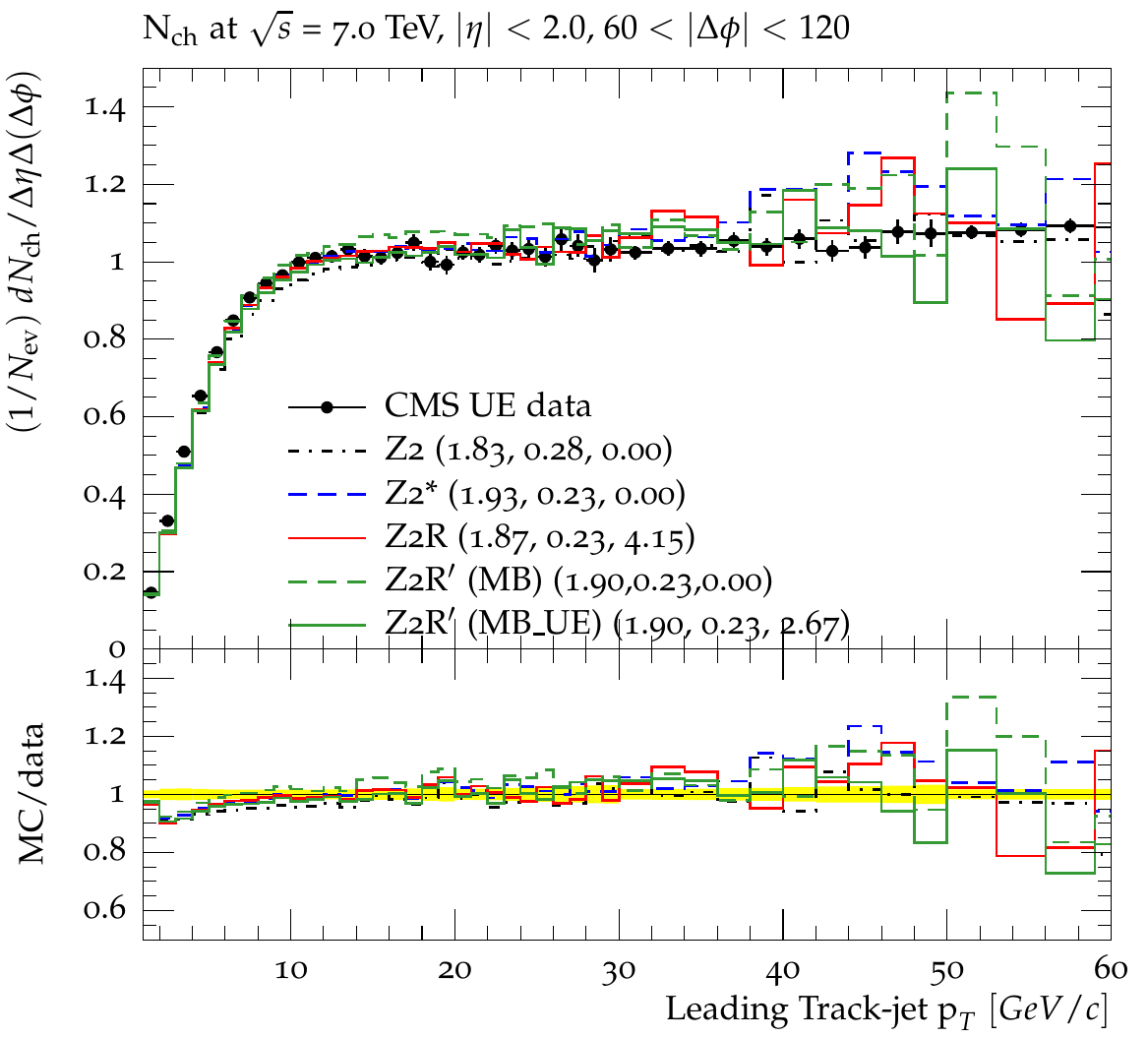}\hfill\mbox{\ }\vskip 0.2em
\hfill\includegraphics[width=0.33\textwidth]{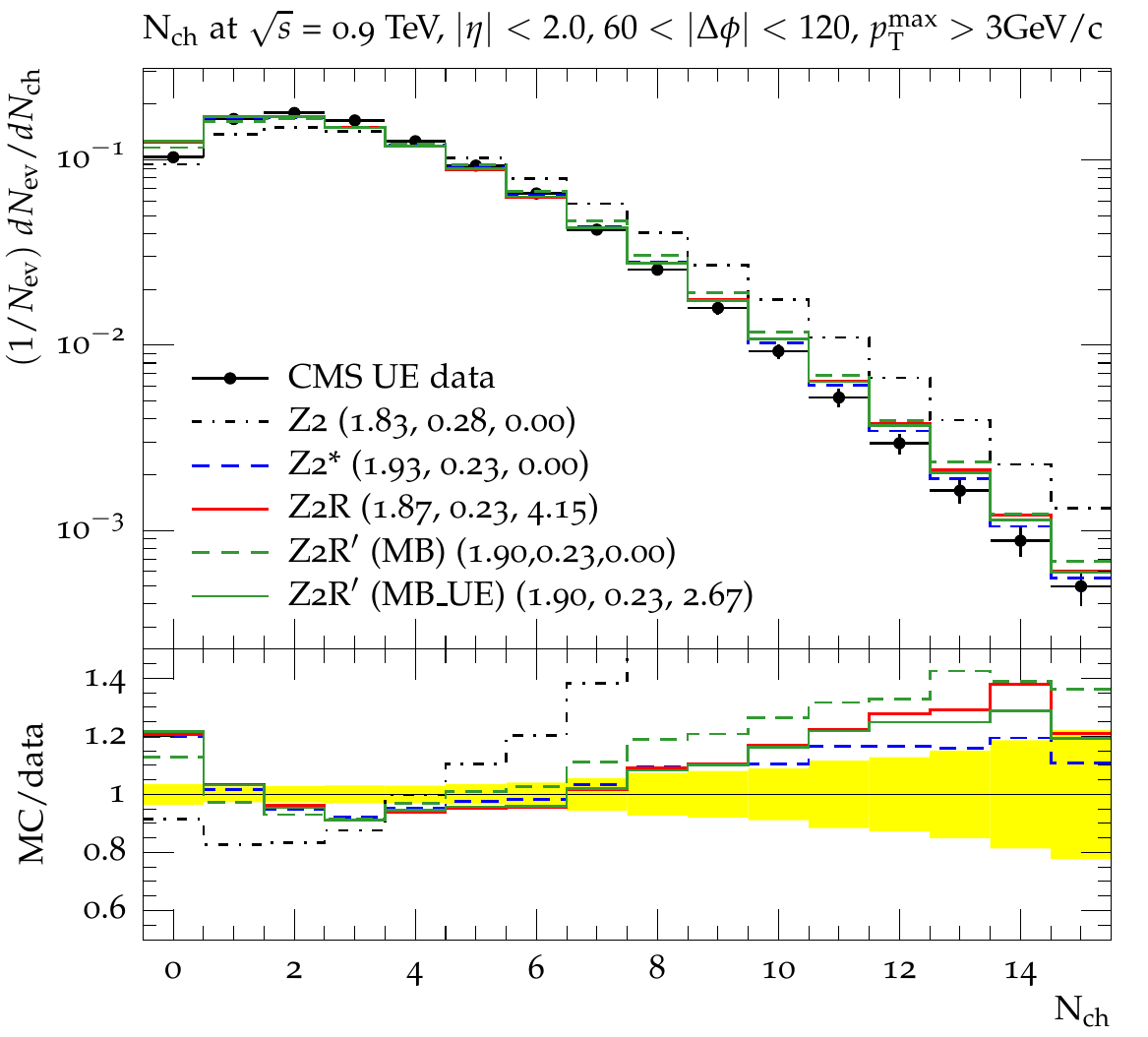}\hspace{1cm}
\includegraphics[width=0.33\textwidth]{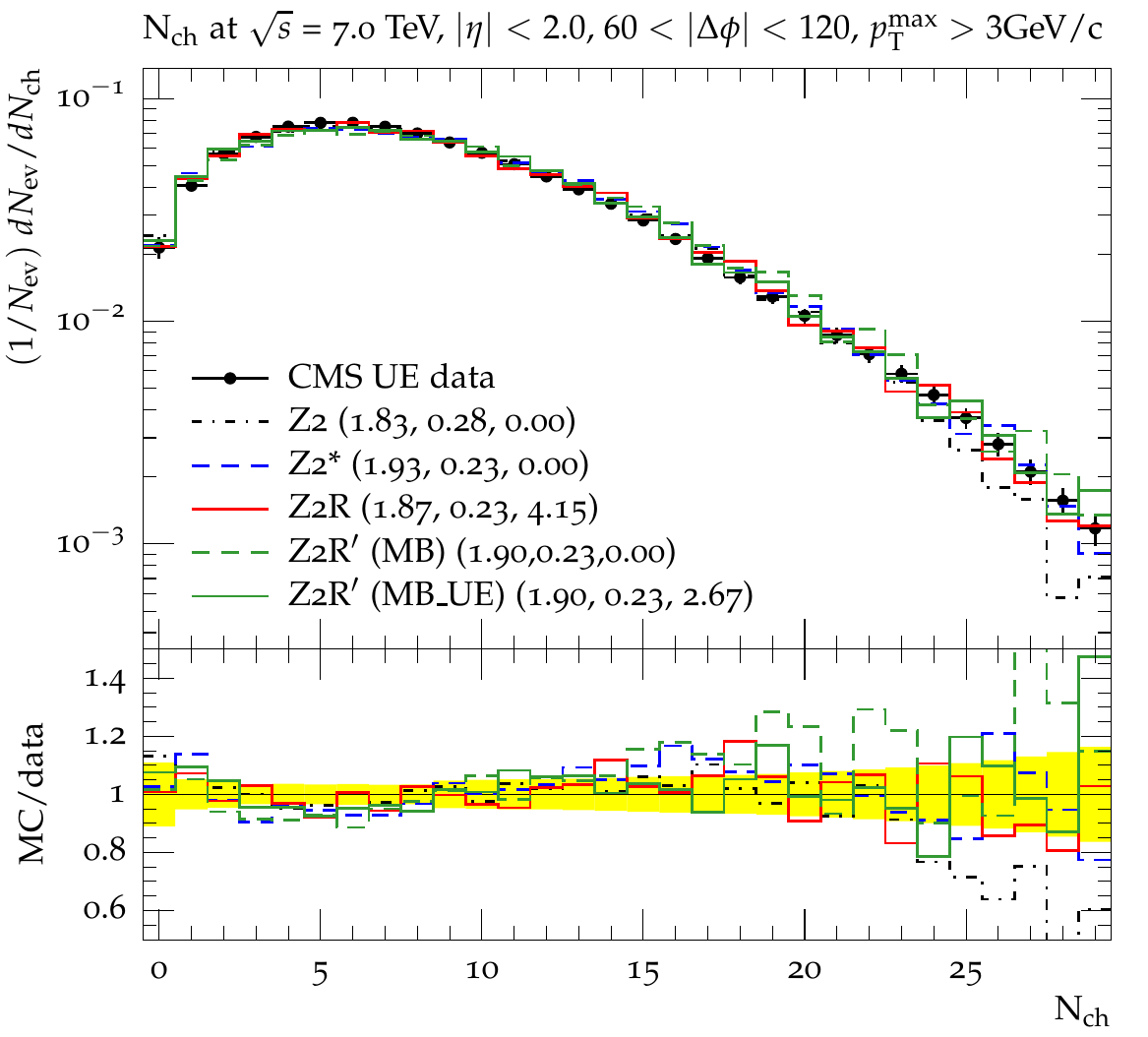}\hfill\mbox{\ }\vskip 0.2em
\hfill\includegraphics[width=0.33\textwidth]{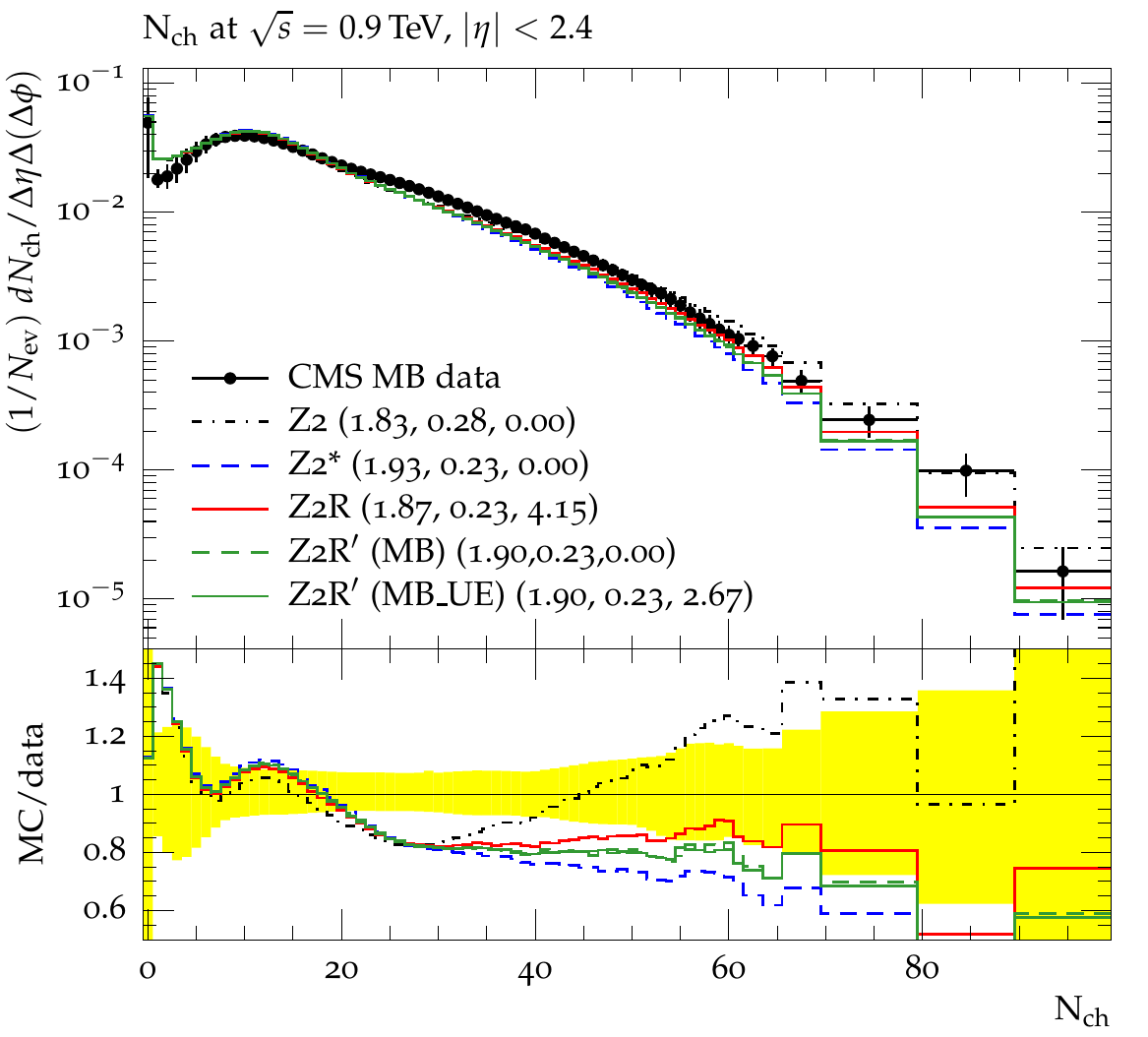}\hspace{1cm}
\includegraphics[width=0.33\textwidth]{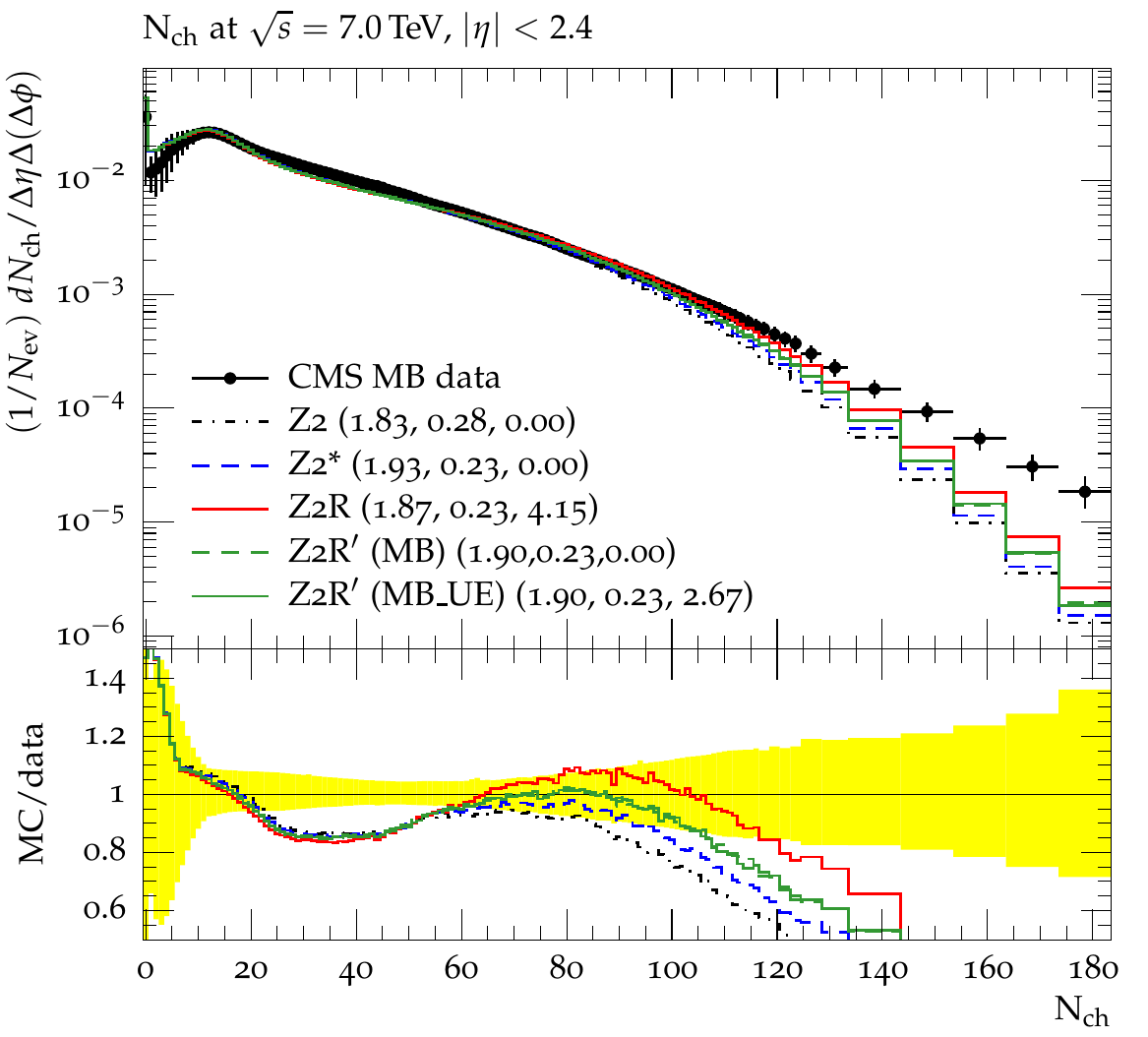}\hfill\mbox{\ }\vskip 0.2em
\hfill\includegraphics[width=0.33\textwidth]{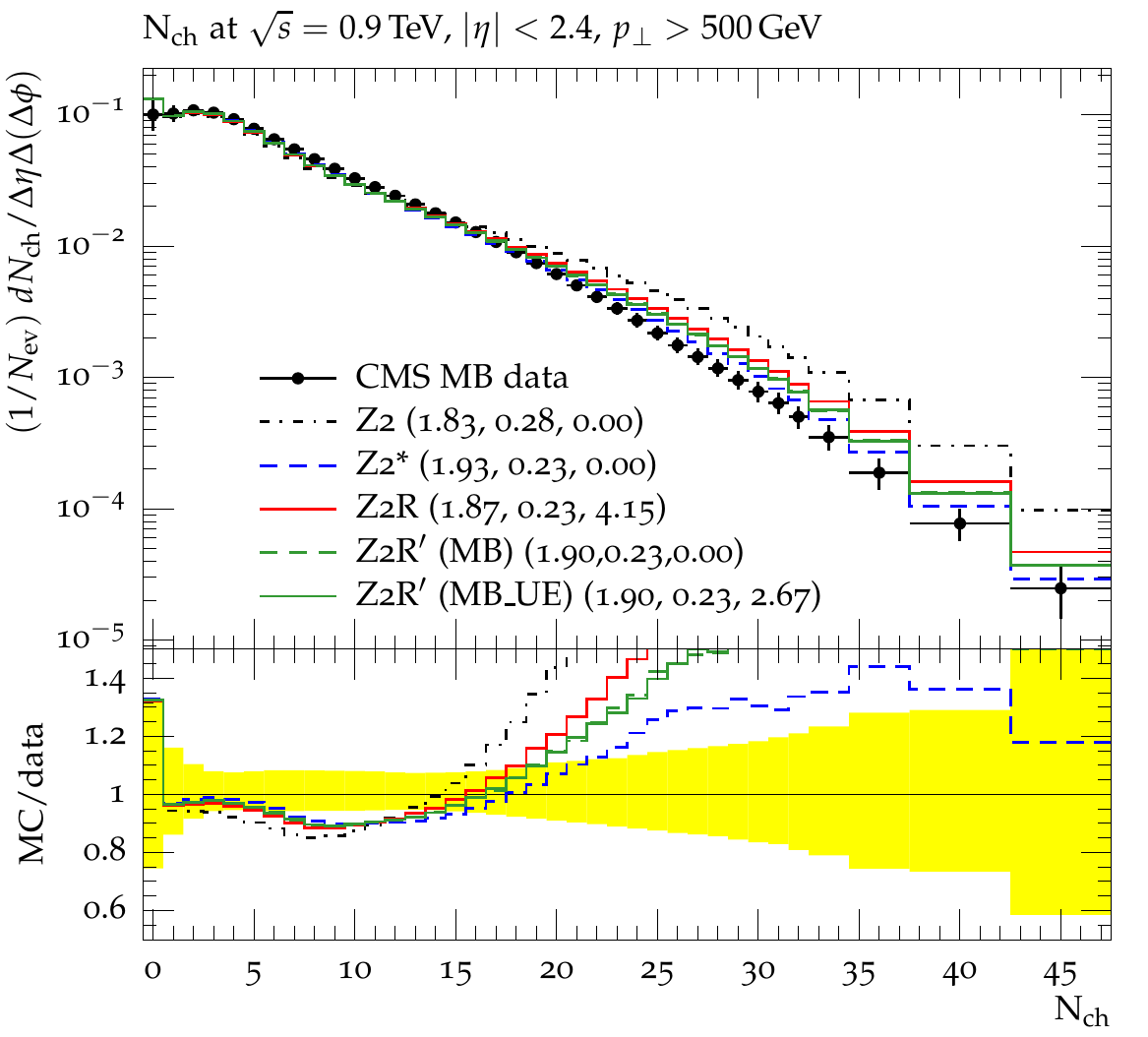}\hspace{1cm}
\includegraphics[width=0.33\textwidth]{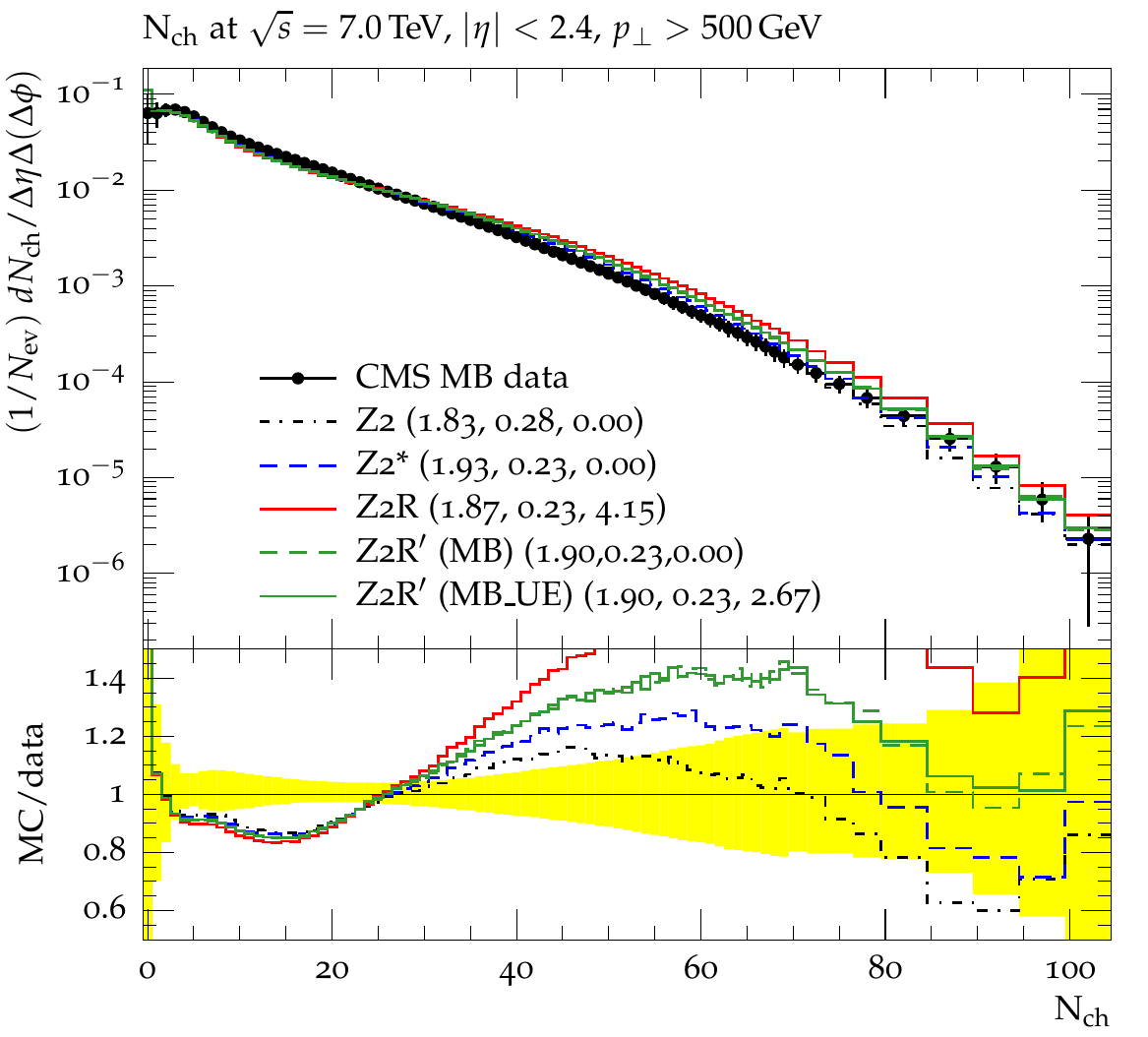}\hfill\mbox{\ }
\definecolor{darkgreen}{rgb}{0.,0.65,0.}
\caption{\label{fig_Z2RZ2Rp}Performance of the \textcolor{red}{Z2R [red solid]} and the \textcolor{darkgreen}{Z2R' [green dashed} (intermediate), \textcolor{darkgreen}{solid} (final)\textcolor{darkgreen}{]} tunes, compared to tunes Z2 [black dashdotted] and \textcolor{blue}{Z2* [blue dashed]}, for select observables in the full UE (rows 1-2) and MB (rows 3-4) data set. The four observables used for tune Z2R are given in rows 1 and 3.}
\end{figure}

The second tune we consider is made in two-steps, we call it Z2R'. This time using all the observables in the same UE and MB CMS datasets, we again perform an automized \textsc{Professor} tune with cubic interpolation. In the first step, the $p_T$-cutoff (both PARP(82) and PARP(90)) is fixed to the MB data, disregarding any match with UE data. In the second step, $\alpha$ is tuned to the UE data. After the first step, the match with data is good for MB, but less so for UE. After the second step, also the match with UE is restored to an acceptable level, comparable to the Z2* tune. Quantitatively, we again find the tune to be insensitive to PARP(90), while PARP(82) and $\alpha$ settle on values inbetween those of Z2* and Z2R (table \ref{tab_Z2Rp}, figure \ref{fig_Z2RZ2Rp}). 

\begin{table}[h]
\centering\footnotesize
\begin{tabular}{|c|c|c|c|}
\hline
& PARP(82) & PARP(90) & $\alpha$ \\ \hline\hline
Z2 & 1.83 & 0.28 & 0.00 \\ \hline
Z2* & 1.93 & 0.23 & 0.00 \\ \hline\hline
Z2R & 1.87 & 0.23 & 4.15 \\ \hline\hline
Z2R' & 1.90 & 0.23 & 2.67 \\ \hline
\end{tabular}
\caption{Result of the 2-step 3-parameter tune to all UE and MB observables.}
\label{tab_Z2Rp}
\end{table}

\subsection{The CMS ridge}
\begin{figure}[!ht]
\centering
\includegraphics[width=0.295\textwidth]{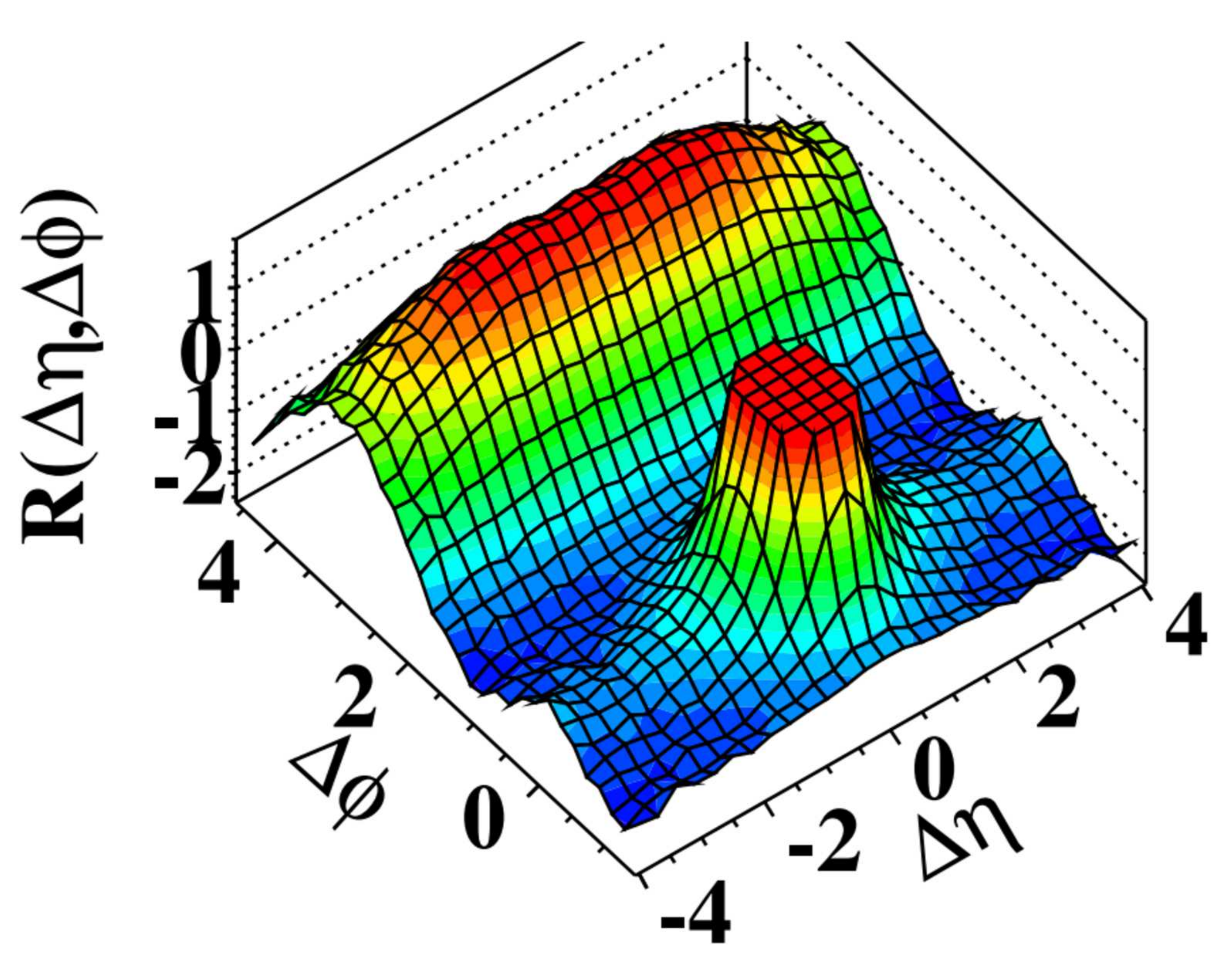}\hfill
\includegraphics[width=0.295\textwidth]{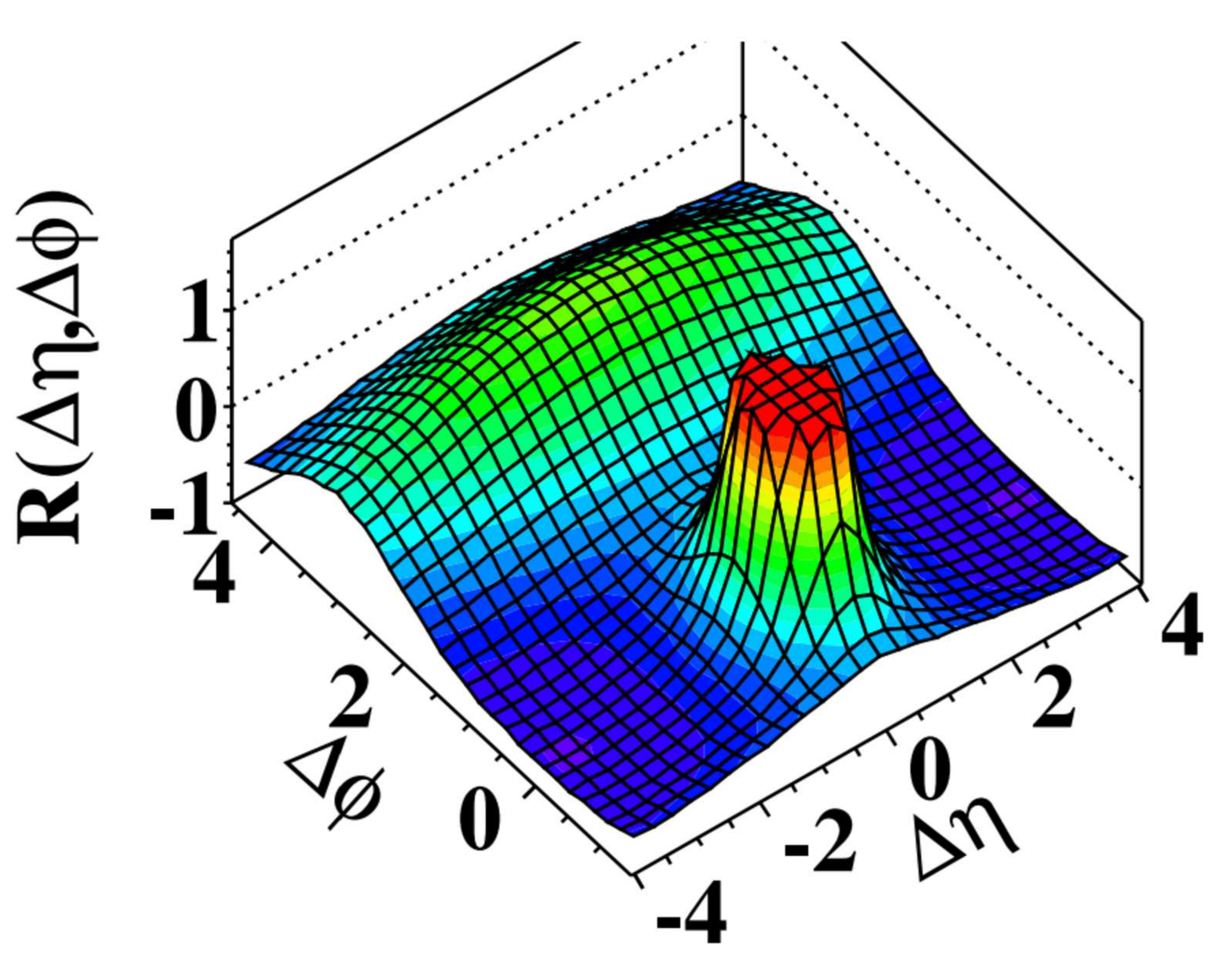}\hfill
\includegraphics[width=0.295\textwidth]{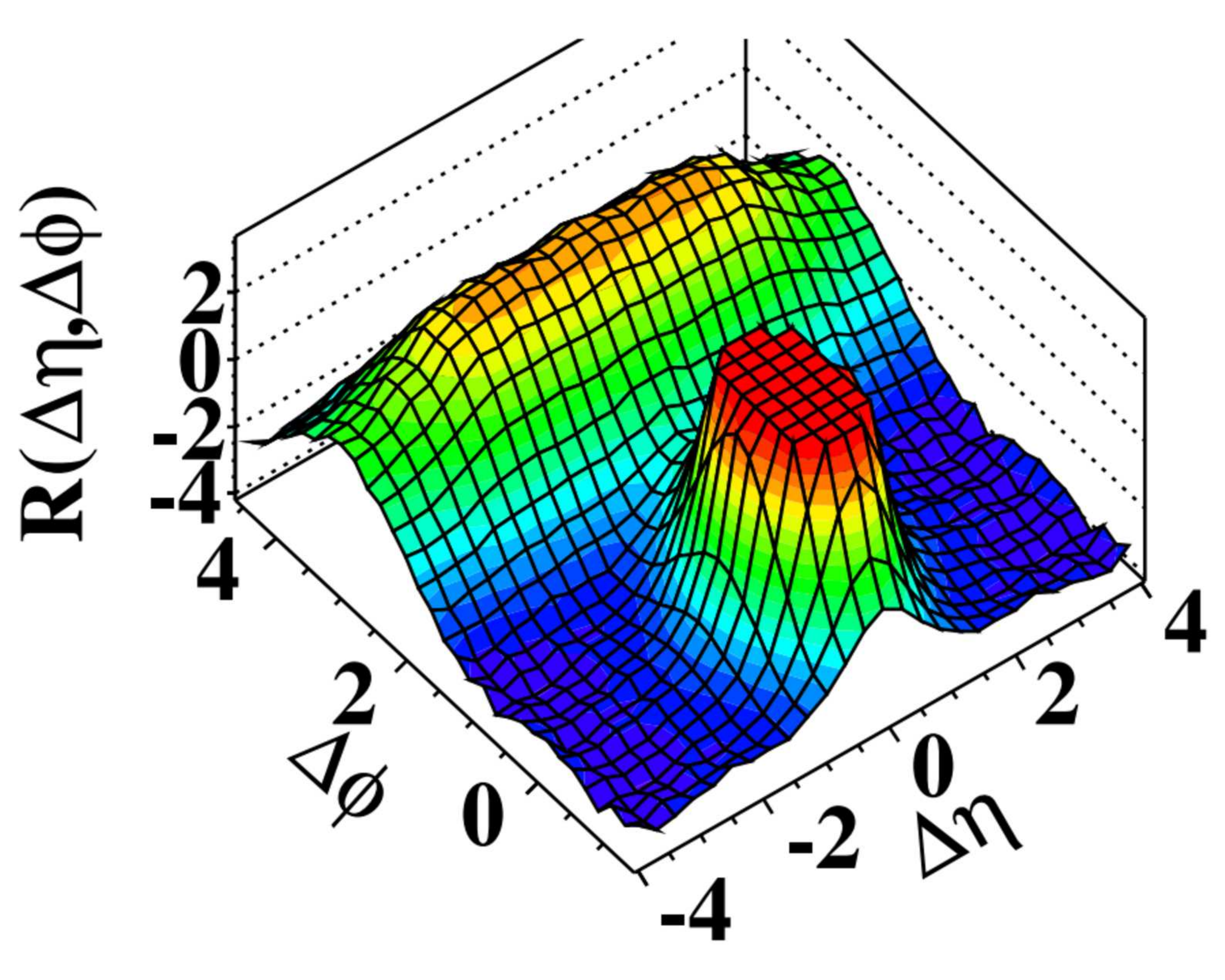}\vskip -0.2em
\includegraphics[width=0.295\textwidth]{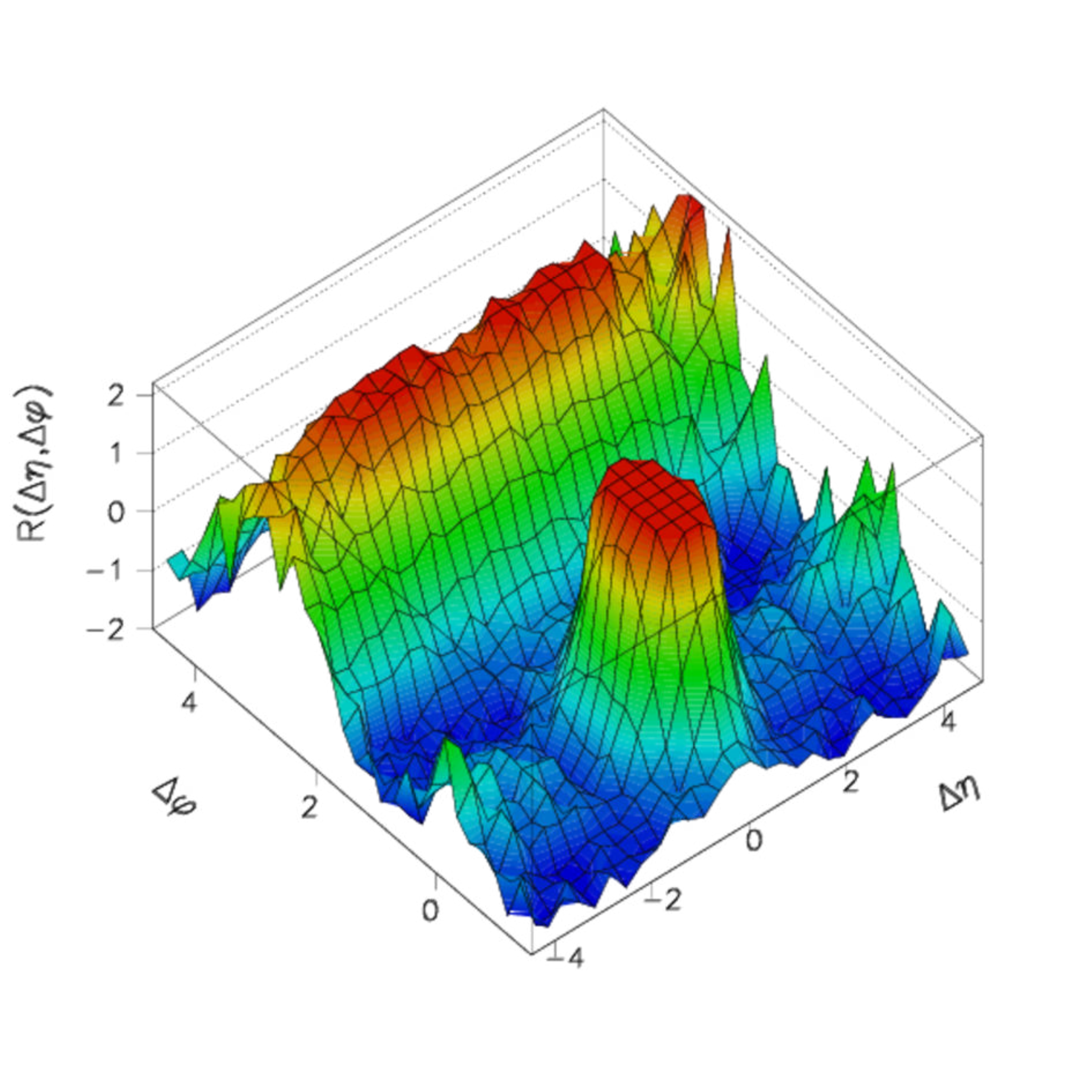}\hfill
\includegraphics[width=0.295\textwidth]{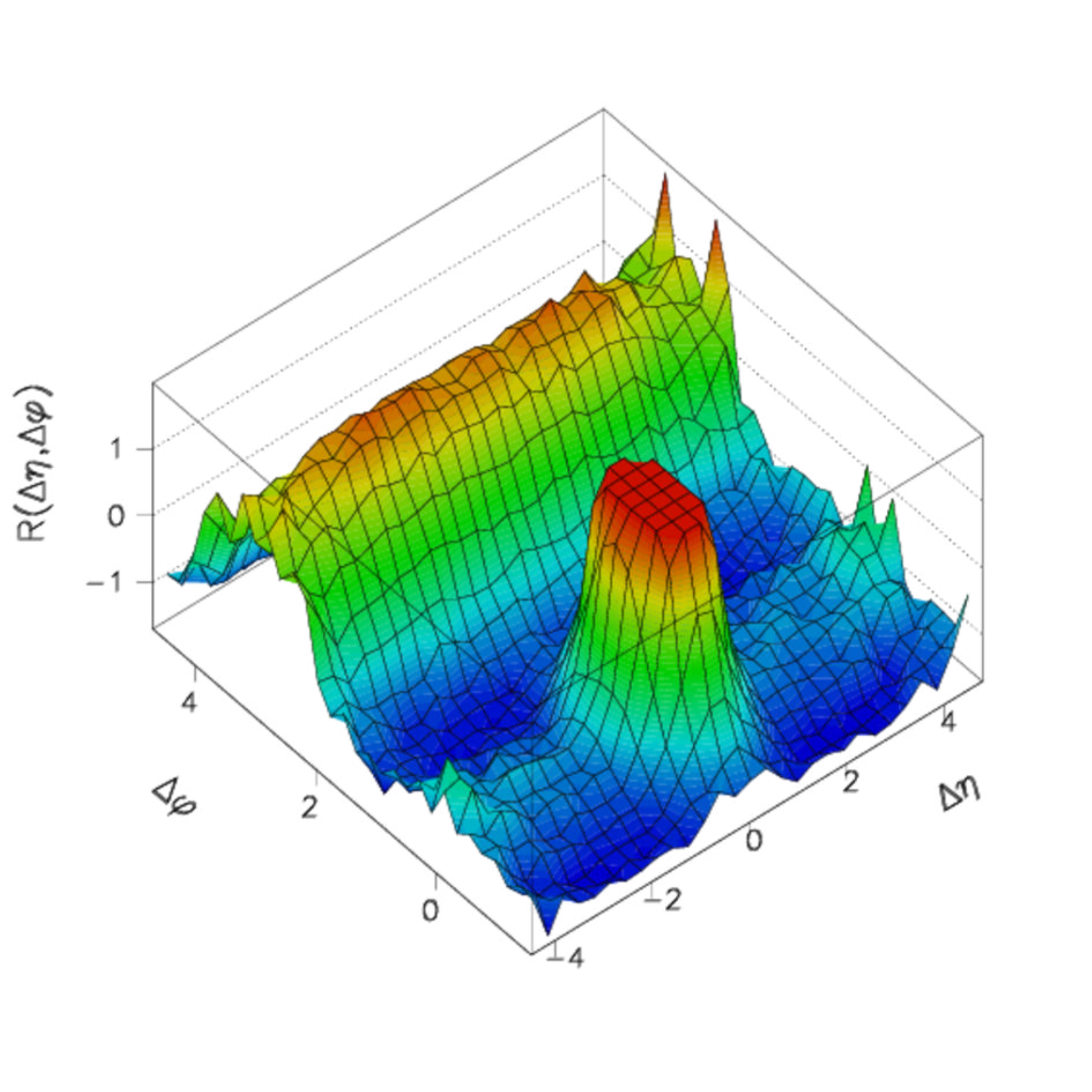}\hfill
\includegraphics[width=0.295\textwidth]{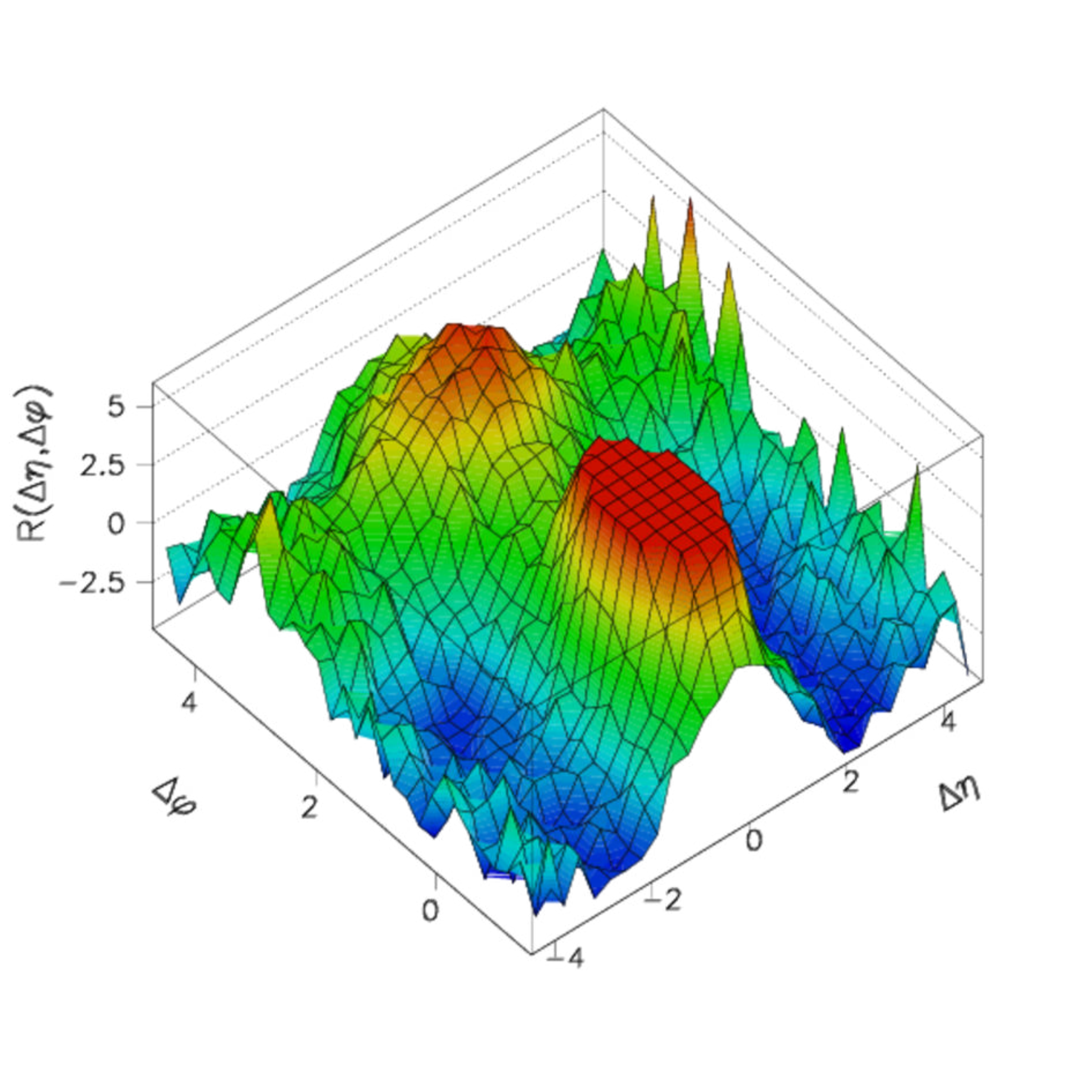}\vskip -0.8em
\includegraphics[width=0.295\textwidth]{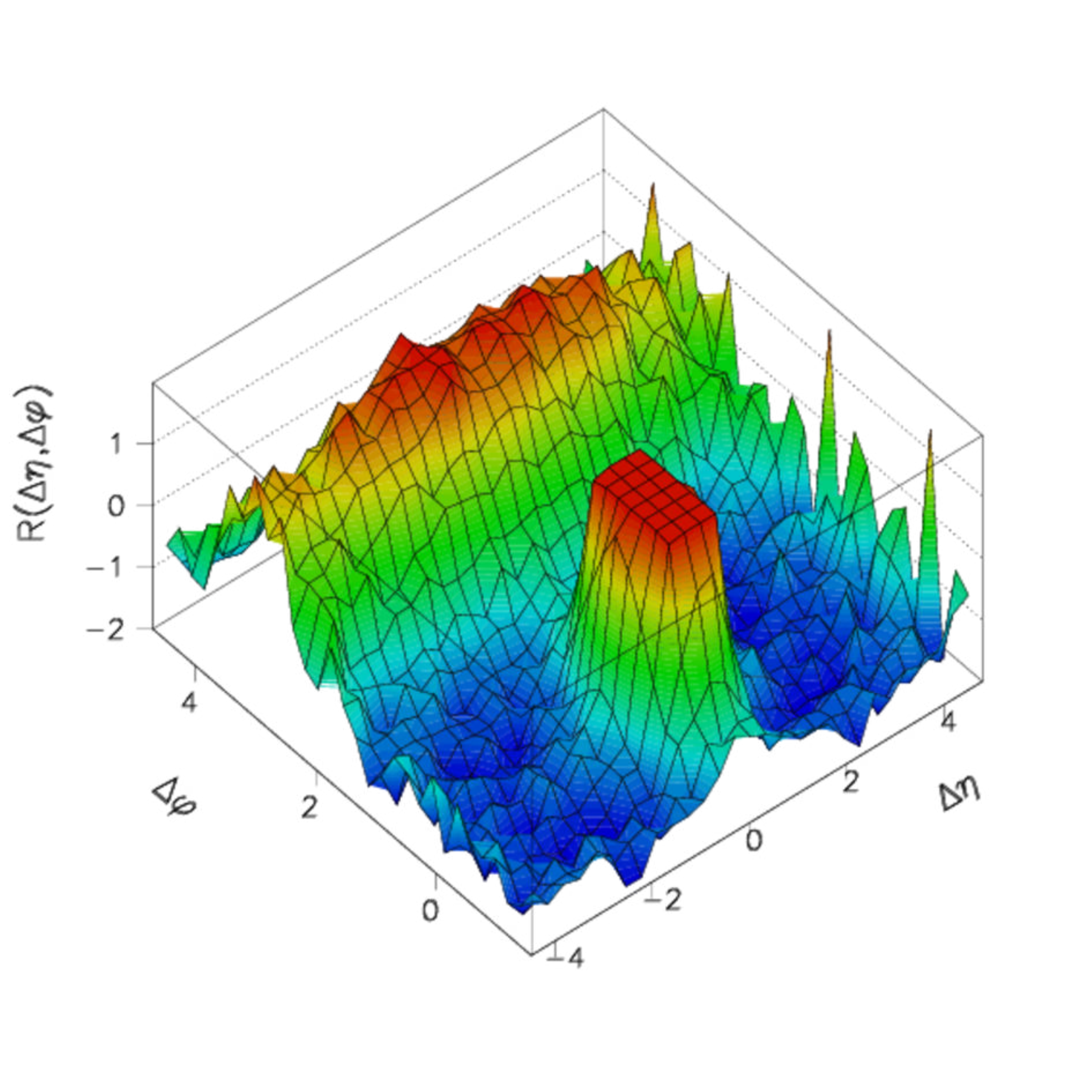}\hfill
\includegraphics[width=0.295\textwidth]{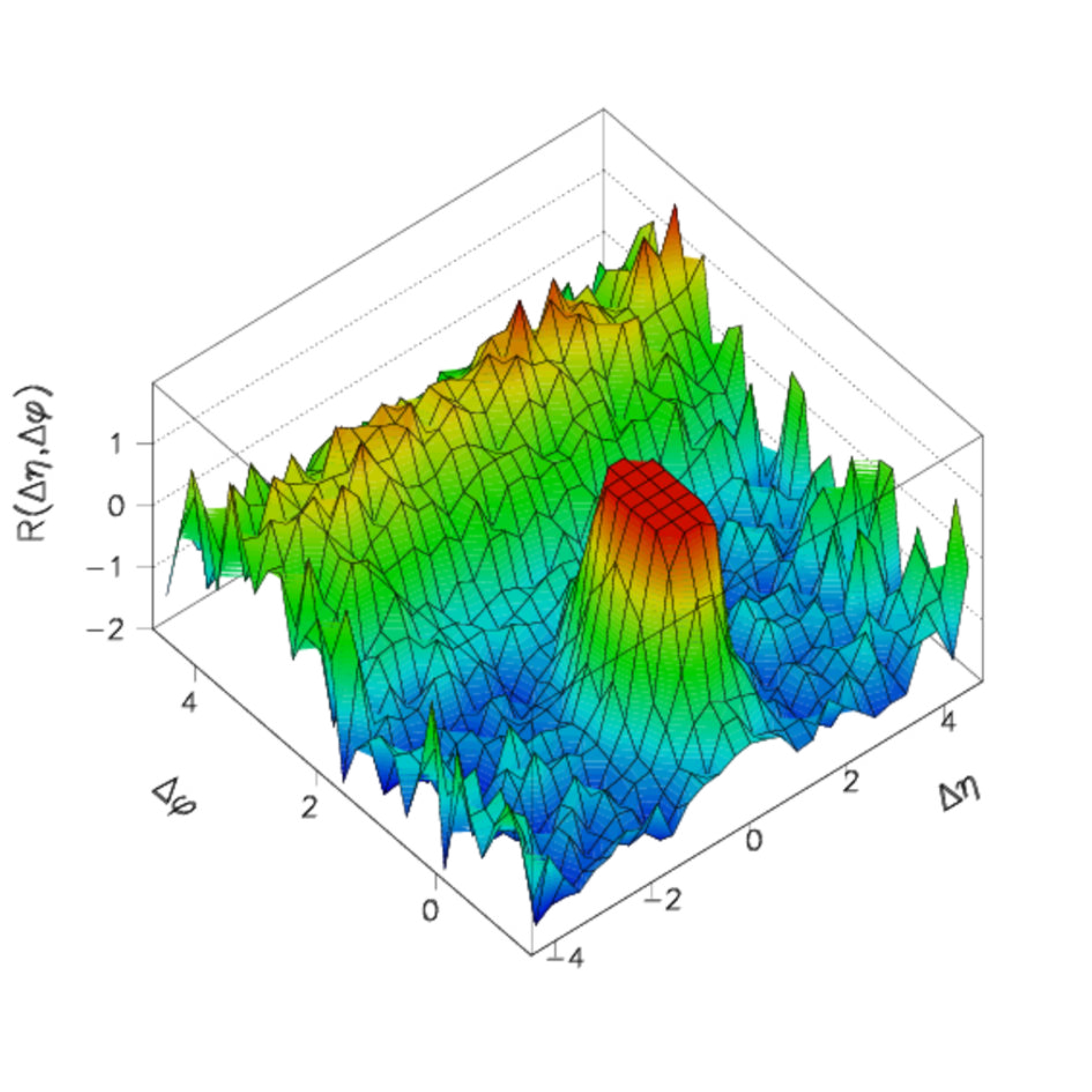}\hfill
\includegraphics[width=0.295\textwidth]{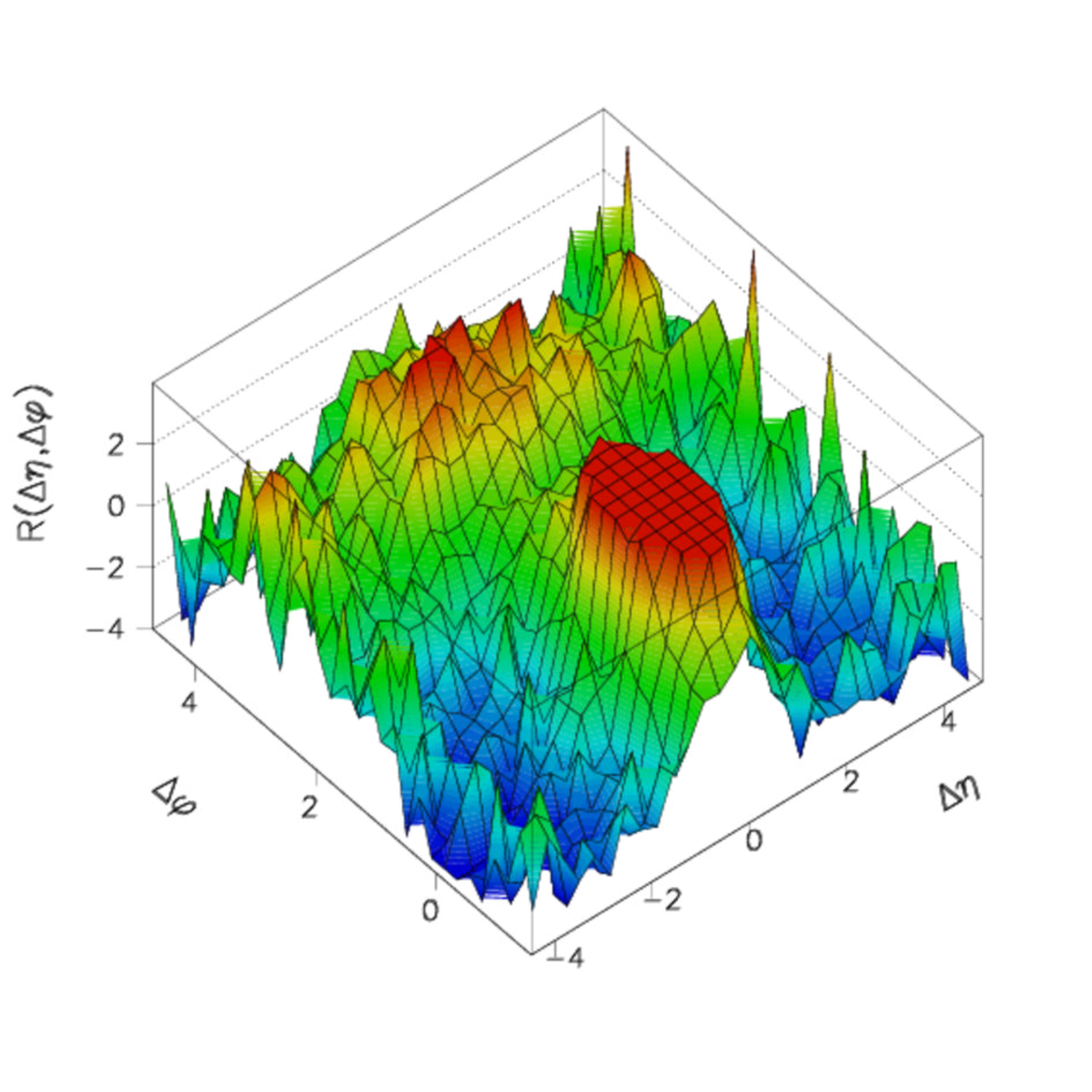}
\caption{\label{ridge}Results for $R\left(\Delta\eta,\Delta\phi\right)$: original (top row), Z2R (middle row) and Z2R' (bottom row); (left) high multiplicity, moderate $p_T$; (centre) minimum bias, moderate $p_T$; (right) high multiplicity, all $p_T > 0.1$~GeV.}
\end{figure}

Finally we consider correlation function $R\left(\Delta\eta,\Delta\phi\right)$, comparing results using tunes Z2R and Z2R' with those from the original paper (figure~\ref{ridge} (top row)). In the middle row, the results for Z2R are shown. It is clear that for high-multiplicity moderate-$p_T$ events (middle, left), the long-range near-side ridge is visible, fully in agreement with the CMS results. In the same row (middle, centre), one can see that also for moderate-multiplicity events a ridge is visible, denoting that perhaps the effect of the modification is too strong. In the bottom row, the same plots are shown for tune Z2R'. Here, the effect is not strong enough at high-multiplicity (bottom, left), as no near-side ridge is visible, while it is still too strong at moderate-multiplicity (bottom, centre), where an unwanted ridge is visible. For high-multiplicity events, including all $p_T > 0.1$~GeV (middle/bottom, right), both tunes show similar effects. There is no near-side ridge and some broadening around $\Delta\eta = 0$ is visible, both in agreement with CMS data, but there is an unexplained additional peak at $\left(\Delta\eta,\Delta\phi\right) = \left(0,\pi\right)$.

\section{Conclusions}
We proposed a modification of \textsc{Pythia6}, explaining the ridge effect with multiparton interactions. The model introduces a correlation between the azimuth of the event planes of individual multiparton interactions and the event plane of the hardest interaction. This correlation can be naturally explained in a physical picture based on the impact parameter between the protons. The two main implications of this modification are the appearance of the near-side ridge in high-multiplicity moderate-$p_T$ events and a shift in the activity in the transverse region. This latter effect can be counteracted by a re-tune of the $p_T$-cutoff parameters to underlying event data. In a slightly broader picture, minimum bias data can be included in the re-tuning. Implementing this with the \textsc{Professor} package, we found tunes Z2R and Z2R'. 

\begin{footnotesize}

\end{footnotesize}

\end{document}